\documentclass{jfm}

\usepackage{graphicx}
\usepackage{natbib}
\usepackage{amsmath}
\usepackage{color}
\usepackage{textcomp}

\ifCUPmtlplainloaded \else
  \checkfont{eurm10}
  \iffontfound
    \IfFileExists{upmath.sty}
      {\typeout{^^JFound AMS Euler Roman fonts on the system,
                   using the 'upmath' package.^^J}%
       \usepackage{upmath}}
      {\typeout{^^JFound AMS Euler Roman fonts on the system, but you
                   dont seem to have the}%
       \typeout{'upmath' package installed. JFM.cls can take advantage
                 of these fonts,^^Jif you use 'upmath' package.^^J}%
      }
  \else
  \fi
\fi

\ifCUPmtlplainloaded \else
  \checkfont{msam10}
  \iffontfound
    \IfFileExists{amssymb.sty}
      {\typeout{^^JFound AMS Symbol fonts on the system, using the
                'amssymb' package.^^J}%
       \usepackage{amssymb}%
       \let\le=\leqslant  
         
      }{}
  \fi
\fi

\ifCUPmtlplainloaded \else
  \IfFileExists{amsbsy.sty}
    {\typeout{^^JFound the 'amsbsy' package on the system, using it.^^J}%
     \usepackage{amsbsy}}
    {\providecommand\boldsymbol[1]{\mbox{\boldmath $##1$}}}
\fi

\providecommand\bnabla{\boldsymbol{\nabla}}

\newcommand\Rey{\mbox{\textit{Re}}}  

\newsavebox{\astrutbox}
\sbox{\astrutbox}{\rule[-5pt]{0pt}{20pt}}

\title[On the need for a nonlinear subscale turbulence term]{On the need for a nonlinear subscale turbulence term 
in POD models
as exemplified for a high Reynolds number flow over an Ahmed body}

\author[Jan \"{O}sth {\em et~al.\/}]%
{Jan \"{O}sth$^1$%
  \thanks{Email address for correspondence: jan.osth@chalmers.se},\ns
 Bernd R. Noack$^2$, Sini\v{s}a Krajnovi\'c$^1$, \break
 Diogo Barros$^{2,3}$ and Jacques Bor\'ee$^2$}

\affiliation{$^1$Division of Fluid Dynamics,
Department of Applied Mechanics,
Chalmers University of Technology, SE-412 96 G\"oteborg, Sweden
\\[\affilskip]
$^2$Institut PPRIME, CNRS - Universit\'e de Poitiers - ENSMA, UPR 3346, D\'epartment Fluides, Thermique, Combustion, CEAT, 43 rue de l'A\'erodrome, F-86036 Poitiers CEDEX, France
\\[\affilskip]
$^3$PSA Peugeot Citro\"{e}n, Centre Technique de V\'elizy, 78943 V\'elizy-Villacoublay Cedex, France}

\pubyear{2010}
\volume{650}
\pagerange{119--126}
\date{?; revised ?; accepted ?. - To be entered by editorial office}
\begin{document}

\maketitle

\begin{abstract}
We investigate a hierarchy of eddy-viscosity terms in POD Galerkin models 
to account for a large fraction of unresolved fluctuation energy.
These Galerkin methods are applied to Large Eddy Simulation data
for a flow around the vehicle-like bluff body called Ahmed body.
This flow has three challenges for any reduced-order model:
a high Reynolds number, coherent structures with broadband frequency dynamics,
and meta-stable asymmetric base flow states.
The Galerkin models are found to be most accurate 
with modal eddy viscosities
as proposed by Rempfer \& Fasel (1994).
Robustness of the model solution
with respect to initial conditions, eddy viscosity values and model order 
is only achieved for state-dependent eddy viscosities
as proposed by Noack, Morzy\'nski \& Tadmor (2011).
Only the POD system with state-dependent modal eddy viscosities 
can address all challenges of the flow characteristics.
All parameters are analytically derived 
from the Navier-Stokes based balance equations with the available data.
We arrive at simple general guidelines 
for robust and accurate POD models
which can be  expected to hold 
for a large class of turbulent flows.
\end{abstract}

\begin{keywords}
\textbf{Nonlinear Dynamics} ---Low-dimensional models;
\textbf{Turbulent Flows} --- Turbulence simulations;
\textbf{Wakes/jets} --- wakes.
\end{keywords}


\section{Introduction}
\label{Sec:Introduction}
In this work, 
we address important enablers for low-dimensional POD models 
for complex high-Reynolds-number flows.
Reduced order models (ROM) are abundantly used in fluid mechanics.
The purposes range from understanding of the physical mechanisms,
to computational inexpensive surrogate models for optimization,
to low-dimensional plants for control design.
In this study, 
we focus on reduced-order Galerkin models,
as they have a convenient mathematical structure 
for the above mentioned purposes.
The Galerkin expansions may arise 
from mathematical completeness considerations \citep{Busse1991asr,Noack1994jfm},
from eigenfunctions of Navier-Stokes related equations \citep{Joseph1976book,Boberg1988zn}
or empirical data \citep{Holmes2012book}.
The majority of low-dimensional Galerkin models
in engineering applications are of empirical nature
and utilize one or another variant
of proper orthogonal decomposition (POD).
The first dynamical POD model was presented 
in the pioneering work of \citet{Aubry1988jfm}. 
Their ROM describes the coherent structures 
in the turbulent boundary layer,
particularly sweeps or ejections.
Other examples are the vortex shedding flow behind a circular cylinder 
at low Reynolds number \citep{Deane1991pfa,Noack2003jfm},
transitional and turbulent boundary layers \citep{Aubry1988jfm,Rempfer1994jfm2},
a turbulent jet and the mixing layer 
\citep{Rajaee1994jfm,Ukeiley2001jfm} 
and the lid-driven cavity flow \citep{Cazemier98pf}. 

POD models have been presented for myriad of flow configurations,
ranging from laminar, to transitional and turbulent states.
Yet, the construction of POD models 
for broad-band turbulence still constitutes a challenge.
A rich set of subscale turbulence representations in POD models have been proposed.
\citet{Aubry1988jfm,Podvin2009pf} employ a single eddy viscosity term,
thus effectively modelling a Navier-Stokes equation at lower Reynolds number.
\citet{Rempfer1994jfm,Rempfer1994jfm2} 
have proposed mode-dependent refinement of eddy viscosities,
inspired by spectral eddy viscosities of homogeneous isotropic turbulence.
All these subscale turbulence representations constitute linear terms
in the mode coefficients.
\citet{Galletti2004jfm} add an additional linear term to the Galerkin system,
calibrating the parameters with a solution matching technique.
Several authors have also proposed nonlinear terms.
\citet{Noack2011book} derive a nonlinear eddy viscosity model,
based on a Finite-Time Thermodynamics (FTT) closure \citep{Noack2008jnet}. 
Nonlinear models based on the Galerkin projection of filtered Navier-Stokes equations 
have been pursued by \citet{Wang2011jcp,Wang2012cmame}.
An approach of completely different nature 
is suggested by \citet{Balajewicz2013jfm}. 
Here, 
no auxiliary subscale turbulence terms have been introduced in the Galerkin system,
but the dissipative effects are incorporated in a generalized POD.

In the present work, 
we present for the first time a ROM for the highly turbulent flow around a three-dimensional vehicle bluff body, 
the so-called Ahmed body.
The Ahmed model is used in vehicle aerodynamics as a generic test case 
that reproduces the important flow structures around passenger vehicles \citep{AhmedSAE1984,DuellSAE199,SpohnAhmed2002,LienhartSAE2003}. 
Recently, the model has been subjected to intensive research for the pursuit of flow control methods capable of reducing the aerodynamic drag on the model, both passive control \citep{BeaudoinEXPF2008,krajnovic:IMECE2013:13}, and active control \citep{BrunnAIAA2006,Pastoor2008jfm,krajnovic:fernandes:10,aider:beaudoin:ef:10}. 
In the present study, we focus on the square-back variant of the Ahmed body, 
which is essentially a bluff body with curved front edges placed in the proximity of ground. 
This flow poses a severe challenge for the ROM due to the bi-modal states of 
the wake that was discovered in the recent study by \citet{Grandemange2013jfm}, 
i.e., the flow switches from one semi-stable asymmetric state to another over time-scales, $T_S$, 
that is of the order of $T_S \approx 100 H/U_{\infty}$ 
where H is the height of the body and $U_{\infty}$ is the velocity of the oncoming flow.

In the proposed POD models,
we employ the modal eddy viscosity refinement by \cite{Rempfer1994jfm2}
and the nonlinear eddy viscosity scaling based on the FTT framework proposed by \citet{Noack2011book} 
to stabilize the long-term solution behaviour. 
The POD model utilizes a dataset of time-resolved flow fields 
of the flow around the bluff body. 
The dataset has been produced by numerical simulations employing the Large Eddy Simulation (LES) technique. The LES data capture the
semi-stable asymmetric states and departures from these states.
The flow around a similar bluff body has been simulated with a LES by \citet{krajnovic:davidson:jfe:02}
over one decade ago. 
The standard \citet{smag:63} subgrid stress model was used both in that study and is used in the present study. 
Although there has been an abundance of more intricate subgrid-stress models developed 
since the days of Smagorinsky half a century ago, his nonlinear model has proven to be robust, 
highly applicable and very capable of producing unsteady solutions to complex bluff body flow cases 
with high accuracy that are able to yield further physical understanding of the flow dynamics. 
For instance, the same LES technique was used to simulate the flow around the Ahmed body with a 25$^{\circ}$ angle of the rear slanted surface by \citet{krajnovic:davidson:jfe:part1:05,krajnovic:davidson:jfe:part2:05}, the flow around high speed trains at low Reynolds numbers by \citet{hemida:noseshape:2008,hemida:noseshape:2010}, 
the flow around freight trains by \citet{hemida:slipstream:2010} and \citet{OesthJFS2014}, 
and the flow around a finite tall circular cylinder by \citet{krajnovic:2011:jfm}.

This manuscript is organized as follows; 
First, the flow configuration with car model
and the Large-Eddy Simulation (LES) 
that was used to produce the dataset of time-resolved flow are presented (\S~\ref{Sec:LES}). 
Next, (\S~\ref{Sec:ROM}) the employed Galerkin models 
with a hierarchy of subscale turbulence representations
are outlined. 
Then, the performance of  
 these POD Galerkin models is studied (\S~\ref{Sec:Results}) and 
conclusions and future directions are provided (\S~\ref{Sec:Conclusions}).

\section{Configuration}
\label{Sec:LES}
This section presents the LES simulation that produced the dataset of flow snapshots serving as input to the empirical Galerkin models.
It begins with the description if the geometry of the vehicle model (\S~\ref{Sec:LES:Model}), followed with the set-up and a brief outline of the LES technique and numerical details of the simulation (\S~\ref{Sec:LES:Configuration}).
The main features of the flow are lastly presented (\S~\ref{Sec:LES:Results}).

\subsection{The Ahmed body model}
\label{Sec:LES:Model}
The employed LES simulation shall reproduce a companion experiment
at Institute PPRIME \citep{Oesth2013tsfp}. \textcolor{black}{A description of these experiments and a comparison of the LES and PIV data is included in appendix \ref{appA}.}
The vehicle model has a square-back geometry. 
The models' length, $L$, is $0.893\,$m, the width, $W$, is $0.35\,$m and the height of the body, $H$, is $0.297\,$m.
All four front edges are rounded with a radius of $r = 0.285\,H$. The model is placed on four cylindrical supports with an oval-shaped cross section 
and the ground clearance, $h$, is $0.168\,H$ ($0.05\,$m). Similar square-back models with curved front were used in the numerical investigation using LES reported in \citet{krajnovic:davidson:jfe:02} and in the joint experimental and numerical study by \citet{verzicco:fatica:aiaaj:2002}. In the present study the Reynolds number based on the height of the model, the free-stream velocity, $U_{\infty}$, and the kinematic viscosity of air at room temperature, $\nu$, is \Rey$_H = 3\cdot10^5$.
\subsection{Flow configuration}
\label{Sec:LES:Configuration}
We consider an incompressible flow of the Ahmed body in a steady finite domain, $\Omega \in R^3$. 
The flow is described in a Cartesian coordinate system $\boldsymbol{x} = (x,y,z)$ 
with unit vectors $\boldsymbol{e}_x,\boldsymbol{e}_y,\boldsymbol{e}_z$, respectively. 
The unit vectors are oriented such that the $x$-direction corresponds to the streamwise direction. 
The $y$-direction corresponds to the wall-normal direction in which the lift force is acting on the bluff body 
and the $z$-direction is aligned with the axis of action of the side force on the body. 
The origin is located at the midpoint of the base face of the Ahmed body.
The plane $z = 0$ thus corresponds to the only symmetry plane of the configuration. 
The time is represented by $t$.
The velocity vector $\boldsymbol{u} = (u,v,w)$,
has $u$, $v$ and $w$ as its  $x$-, $y$- and $z$-components, respectively.
The pressure field is denoted by $p$.
In the following, all quantities are normalized with respect
to the oncoming velocity $U_{\infty}$, the Ahmed body height $H$
and the constant density $\rho$ of the fluid.
The flow is described by the incompressible Navier-Stokes Equation (NSE) 
with corresponding initial and boundary conditions, $\boldsymbol{u}_{IC}$ and $\boldsymbol{u}_{BC}$, respectively:
\begin{subeqnarray}
 \label{eq:gov_equationsPOD}
\partial _t \boldsymbol{u} + \boldsymbol{u} \cdot \bnabla \boldsymbol{u} + \bnabla p - \nu \nabla ^2  \boldsymbol{u} &=& 0, \\[3pt]
\bnabla \cdot  \boldsymbol{u} &=& 0, \\[3pt]
\boldsymbol{u}(\boldsymbol{x},0) &=& \boldsymbol{u}_{IC}(\boldsymbol{x})\,\,\, \forall \boldsymbol{x} \in \Omega, \\[3pt]
\boldsymbol{u}(\boldsymbol{x},t) &=& \boldsymbol{u}_{BC}(\boldsymbol{x})\,\,\, \forall \boldsymbol{x} \in \partial \Omega, t \in [0,T].
\end{subeqnarray}

Here, $\nu=1/Re_H$ represents the non-dimensionalised kinematic viscosity, 
or, equivalently, the reciprocal Reynolds number.
The length of the investigated  time interval  $[0,T]$ is $T =$ \textcolor{black}{500} time units,
after the flow has converged to its post-transient time.
For later reference, 
we define the residual of the momentum equation,
\begin{equation}
\label{eqn:residual}
\boldsymbol{R}(\boldsymbol{u}) = \partial _t \boldsymbol{u} + \boldsymbol{u} \cdot \bnabla \boldsymbol{u} + \bnabla p - \nu \nabla ^2  \boldsymbol{u}.
\end{equation}

This residual is considered as function of the velocity field,
since the pressure can be computed from the velocity field by the pressure-Poisson equation.

\subsection{Large Eddy Simulation (LES)}
\label{Sec:confles}
\begin{figure}
  \centerline{\includegraphics[width=130mm,trim=0cm 10cm 0cm 9cm,clip=true,keepaspectratio]{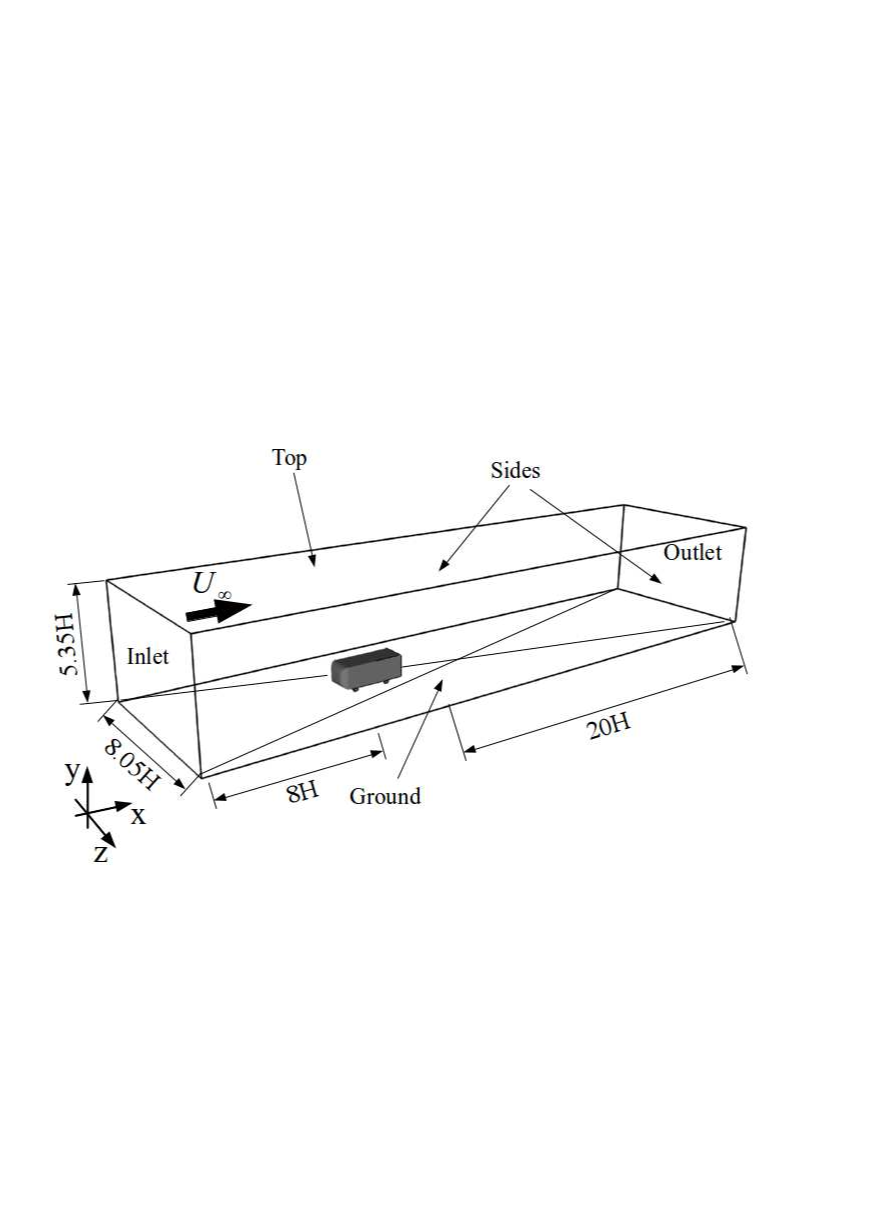}}
  \caption{The computational domain used in the LES simulation.}
\label{fig:1}
\end{figure}
The database of the time-resolved flow around the Ahmed model 
that serves as the input to the empirical Reduced Order Model 
was produced using numerical simulations employing the LES technique.
The governing filtered incompressible Navier-Stokes Equations are closed using the non-linear subgrid-stress model originally proposed by \citet{smag:63}. 
The method has already been used in numerous scientific investigations of vehicle aerodynamics bluff body flows \citep[see e.g.][]{krajnovic:phdthesis:2002,hemida:phdthesis:2008,krajnovic:brs:09,THIELE:LESAHMED:2009,OSTHKRAJNOVIC:TRACTORTRAILER:JWEIA}. The LES equations are discretised using a commercial finite-volume code \citep{AVL2013} using a co-located grid arrangement and the discretised equations are solved for the velocities. The pressure is obtained by a pressure-correction procedure. The employed computational grid consists of 34 million grid points and the obtained spatial resolution was fine enough to be considered a well-resolved LES according to the common conventions in the field \citep{howtoestimate:davidson:2010}. \textcolor{black}{The spatial resolution is detailed in appendix \ref{appB}}. The convective fluxes are approximated by a blend of $95\,\%$ linear interpolation of second order accuracy (Central Differencing Scheme) and of $5\,\%$ upwind differences of first order accuracy (Upwind 
Scheme). The diffusive terms containing viscous plus sub-grid 
terms are approximated by a central differencing interpolation of 
second order accuracy. The time marching procedure is done using the implicit second-order accurate three-time level scheme.
The computational domain is shown in figure \ref{fig:1}. On the inlet a uniform velocity profile in the streamwise direction ($x$-direction) is applied with the freestream velocity $U_{\infty}$.   
On the outlet the homogeneous Neumann condition is used and on the sides the symmetry condition is used. On the ground a slip condition is set on the first part from the inlet to the body in order to prevent the boundary layer development here. On the rest of the ground the no-slip condition is enforced. \textcolor{black}{This no slip condition on the ground is set in order to match the experimental set-up, where the model is mounted on a plate above the ground (see appendix \ref{appA}).}

\subsection{Flow characteristics}\label{Sec:LES:Results}
\begin{figure}
\centerline{(\textit{a})\includegraphics[height=50mm,trim=0cm 0cm 0cm 0cm,clip=true,keepaspectratio]{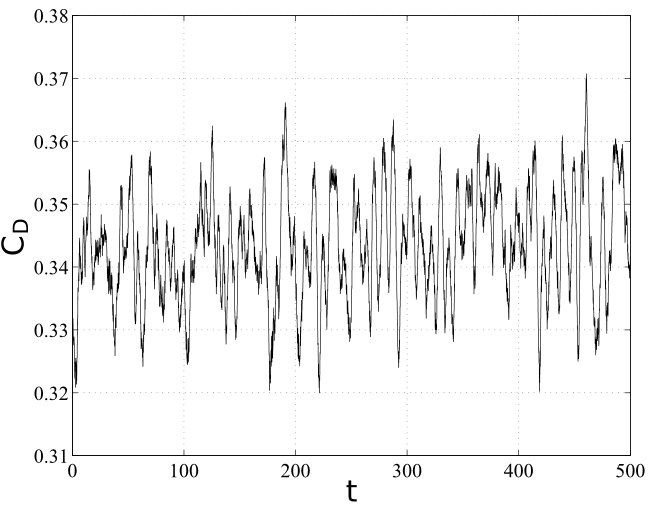}
(\textit{b})\includegraphics[height=50mm,trim=0cm 0cm 0cm 0cm,clip=true,keepaspectratio]{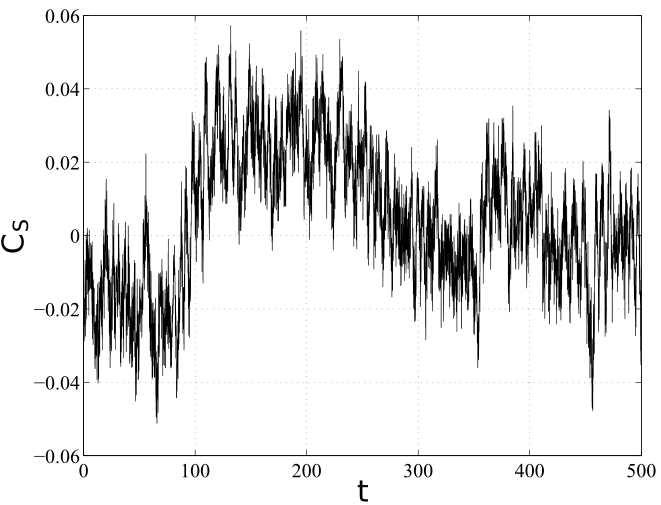}}
\caption{Time history of the force signals from the LES simulation of the natural flow:
(\textit{a}) drag force; 
(\textit{b}) side force.}
\label{fig:2}
\end{figure}

Figure \ref{fig:2}(\textit{a}) presents the time history of the normalized drag force signal, $C_D$, from the simulation. 
A spectral analysis of the signal reveals several low frequency peaks at Strouhal numbers $St = t \cdot H/U_{\infty}$ = 0.036, 0.054, 0.085, 0.12, 0.17 and 0.21, but no dominating peak is found, indicating a broad band spectrum of the flow structures in the wake. Figure \ref{fig:2}(\textit{b}) shows the side force signal, $C_S$. The aerodynamic coefficients are defined as

\begin{eqnarray}
C_D = \frac{F_x}{\frac{1}{2} \rho U_{\infty}^2 A_x} \;\;  ; \;\; C_S = \frac{F_z}{\frac{1}{2} \rho U_{\infty}^2 A_x}.
\end{eqnarray}

Here, $F_x$ and $F_z$ is the total force (pressure and viscous) integrated over the body in streamwise and transversal direction, respectively. $A_x = H \cdot W$ is the cross-sectional area of the Ahmed body. 

The switch between one bi-modal state to the other is clearly indicated in the figure \ref{fig:2}(\textit{b}). The time interval used to plot the forces in figure \ref{fig:2} corresponds to the time-domain that is covered by the snapshots used for the POD and the Galerkin models.

\begin{figure}
\centerline{(\textit{a})\includegraphics[width=50mm,trim=1cm 0cm 1cm 0cm,clip=true,keepaspectratio]{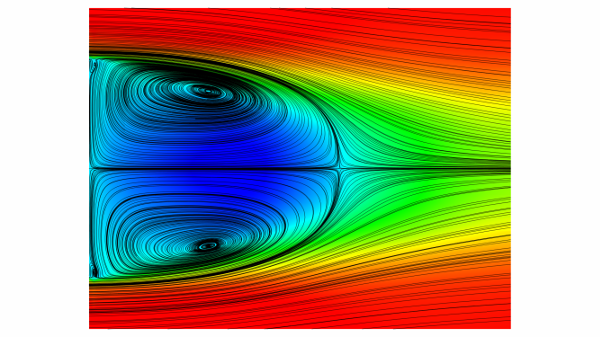} (\textit{b})\includegraphics[width=50mm,trim=1cm 0cm 1cm 0cm,clip=true,keepaspectratio]{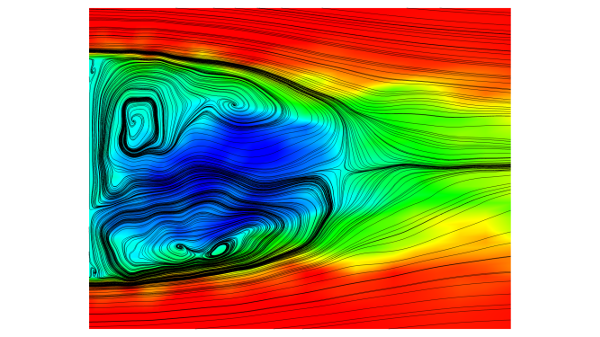} \includegraphics[width=20mm,trim=7cm 10cm 7cm 10cm,clip=true,keepaspectratio]{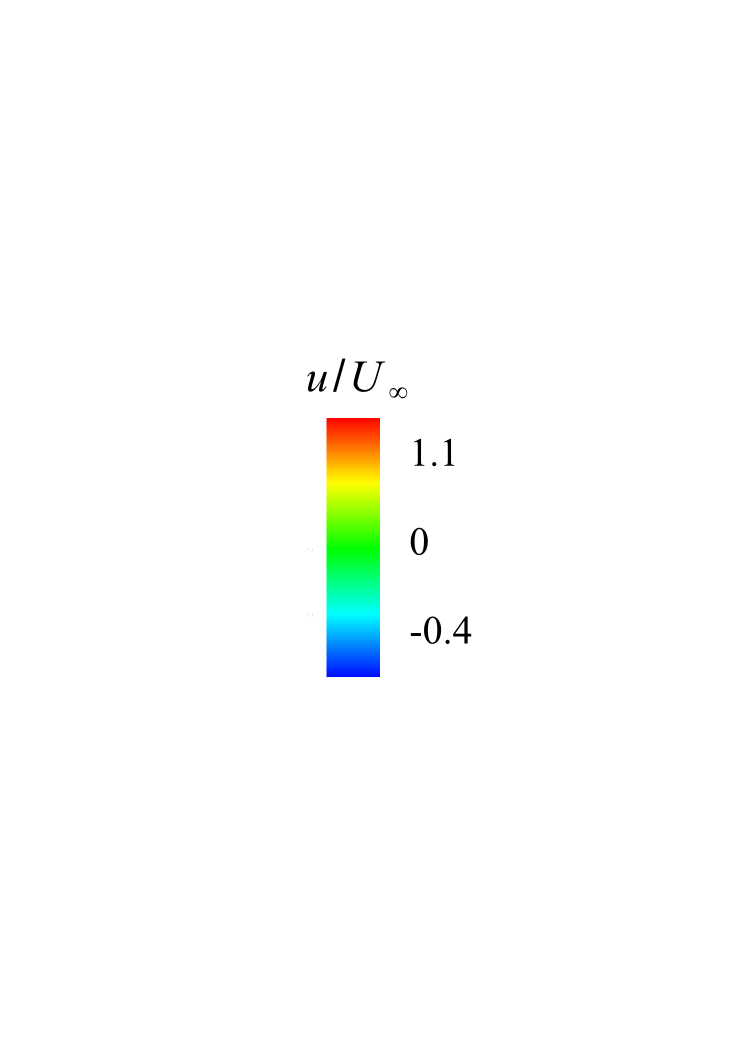} }
\centerline{(\textit{c}) \includegraphics[width=50mm,trim=1cm 0cm 1cm 0cm,clip=true,keepaspectratio]{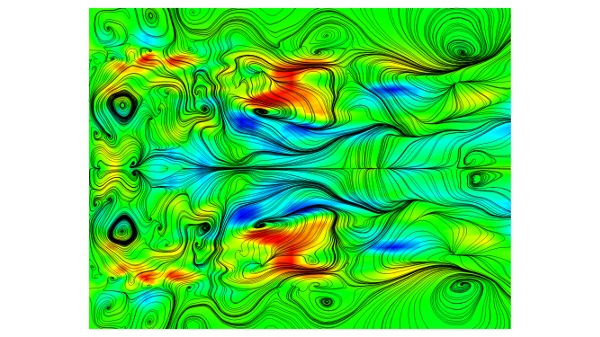} (\textit{d})\includegraphics[width=50mm,trim=1cm 0cm 1cm 0cm,clip=true,keepaspectratio]{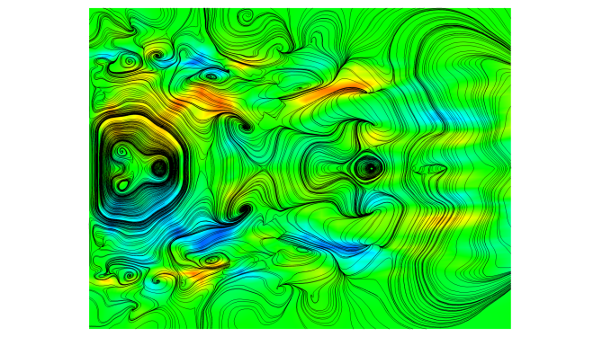} \includegraphics[width=20mm,trim=7cm 10cm 7cm 10cm,clip=true,keepaspectratio]{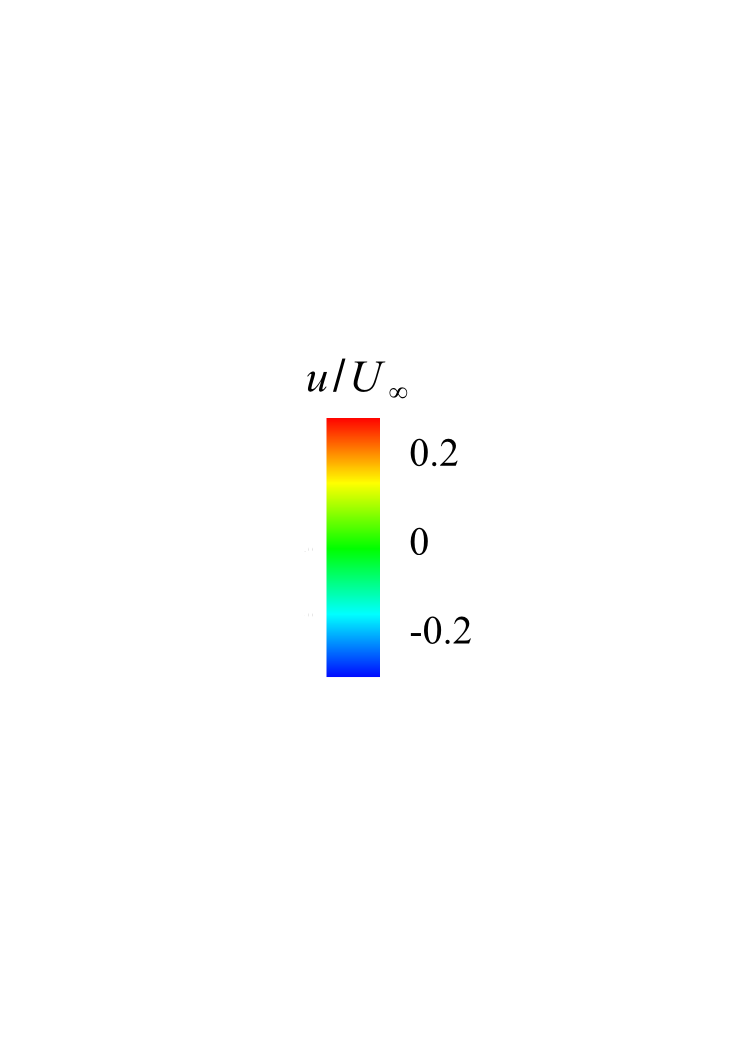} }
\centerline{(\textit{e})\includegraphics[width=70mm,trim=2cm 10cm 2cm 10cm,clip=true,keepaspectratio]{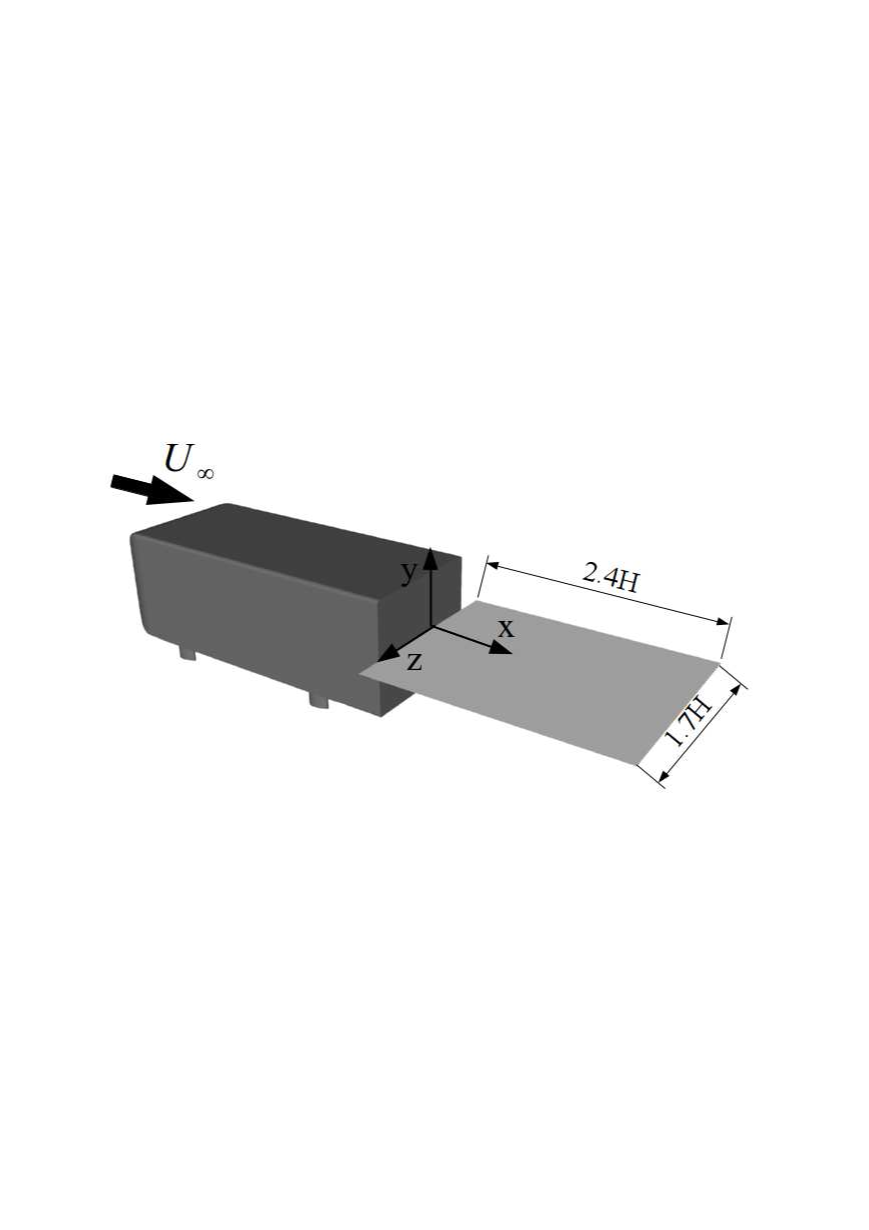} }
\caption{Visualizations of the wake flow:
(\textit{a}) time-averaged symmetrized flow; 
(\textit{b}) one instantaneous realization; 
(\textit{c}) the symmetric part of the instantaneous realization;
(\textit{d}) the anti-symmetric part of the instantaneous realization;
(\textit{e}) the Ahmed model and the plane used to visualize the flow.}
\label{fig:3}
\end{figure}

Figure \ref{fig:3} shows a small selection of the flow results in the wake in the plane $y = 0$ (pointed out in figure \ref{fig:3}(\textit{e})) to help the reader to get an appreciation of the flow behaviour in the wake. In the POD we have used the method described by \citet{Sirovich1987qam2} to split the original data into two sets: one set that is symmetric with respect to the symmetry plane($z = 0$), and one which is anti-symmetric. This procedure will be described in detail in \S~\ref{Sec:ROM:Snapshots}.

Figure \ref{fig:3}(\textit{a}) shows the symmetrized mean flow, $\boldsymbol{u}(\boldsymbol{x})$. Figure \ref{fig:3}(\textit{b}) shows one instantaneous realization of the flow and figures \ref{fig:3}(\textit{c}) and \ref{fig:3}(\textit{d}) show the corresponding symmetric and anti-symmetric decomposition of that snapshot, respectively. Here, the mean flow has been subtracted from the symmetric snapshot (the anti-symmetric mean is zero) so that it corresponds to the input of the POD.

\section{Reduced Order Modelling}
\label{Sec:ROM}
In this section, the path to the POD model is outlined. 
First in \S~\ref{Sec:ROM:Snapshots}, 
the employed LES data and its symmetrisation is outlined.
Then the POD expansion is described in \S~\ref{Sec:ROM:POD}.
Finally in \S~\ref{Sec:ROM:GS},
the refined subscale turbulence representations are discussed.

\subsection{LES snapshots}
\label{Sec:ROM:Snapshots}
The POD is based on $M=1000$ snapshots of the LES.
\textcolor{black}{The sampling frequency is 2, i.e.\ 500 convective time units are covered.}
The convective  time unit is based on $H$ and $U_{\infty}$.
The statistical symmetry with respect to the $z=0$ plane
is enforced following \citet{Sirovich1987qam2}.
This symmetrisation increases the accuracy of the POD decomposition.

Each velocity field 
is decomposed into a symmetric and antisymmetric contribution
with respect of the plane $z = 0$,
$$\boldsymbol{u}(x,y,z) = \boldsymbol{u}_s(x,y,z) + \boldsymbol{u}_{as}(x,y,z).$$
Here, the symmetric part $\boldsymbol{u}_s$ is defined by
\begin{subeqnarray}
 \label{eq:us}
 u_{s}(x,y,z) = \frac{1}{2}(u(x,y,z) + u(x,y,-z)), \\
 v_{s}(x,y,z) = \frac{1}{2}(v(x,y,z) + v(x,y,-z)), \\
 w_{s}(x,y,z) = \frac{1}{2}(w(x,y,z) - w(x,y,-z)).
\end{subeqnarray}
while the anti-symmetric component $\boldsymbol{u}_{as}$  reads
\begin{subeqnarray}
 \label{eq:uas}
 u_{as}(x,y,z) = \frac{1}{2}(u(x,y,z) - u(x,y,-z)), \\
 v_{as}(x,y,z) = \frac{1}{2}(v(x,y,z) - v(x,y,-z)), \\
 w_{as}(x,y,z) = \frac{1}{2}(w(x,y,z) + w(x,y,-z)).
\end{subeqnarray}
Thus, $M = 1000$ snapshots create equal amounts of symmetric and anti-symmetric snapshots. 
The POD is performed on each of the symmetrized sets separately.
The resulting two POD  are combined in a single POD 
and sorted according to their energy level. 
Thus, we have in total 2000 POD modes.
This procedure has already been recommended by \citet{Sirovich1987qam2}
and guarantees the expected statistical symmetries
of the snapshot ensemble.
In addition, the POD modes are either symmetric or anti-symmetric
as derivable from theory.

In principle, 
the same results can be achieved by a simpler method:
the more commonly employed inclusion
of $M$ mirror-symmetric snapshots in the snapshot ensemble.
However, in practice,
pure symmetric and anti-symmetric POD modes
are only guaranteed in their corresponding subspace.
We observed that some symmetric and anti-symmetric POD modes 
with very similar energy levels in the first approach
yield  2 non-symmetric (mixed) modes in the second approach
due to numerical errors.

\subsection{Proper Orthogonal Decomposition}
\label{Sec:ROM:POD}
We perform a Proper Orthogonal Decomposition (POD) expansion \citep{Lumley1970book} 
of $M$ temporally equidistantly sampled velocity snapshots 
$\boldsymbol{u}^m := \boldsymbol{u}(\boldsymbol{x},t^m)$
at times $t^m = m \Delta t$, $m=1,\ldots,M$
with the time step $\Delta t$.
The averaging operation of any velocity dependent function $\boldsymbol{F} ( \boldsymbol{u} )$ 
over this ensemble is denoted by angular brackets,
\begin{equation}
\label{eq:aveoperator}
\langle \boldsymbol{F} ( \boldsymbol{u} ) \rangle 
:= \frac{1}{M} \sum_{m=1}^M \boldsymbol{F} \left( \boldsymbol{u}^m \right).
\end{equation}
The colon in front of the sign emphasizes that the left-hand side
is defined by the right-hand side of the equation.
The observation region $\Omega_{ROM} \subset \Omega$
is a wake centred subset of the computational domain
\begin{equation}
\label{eq:reduceddomain}
\Omega_{ROM} = \left\{ 
(x,y,z) \in \Omega \> \colon \> 0\le x \le 5\,H, -0.67\,H \le y \le 1.12\,H, \vert z \vert \le 1.21\,H.
\right\}
\end{equation} 
This domain is large enough to resolve the recirculation region 
and the absolutely unstable wake dynamics
but small enough to keep the model dimension affordable. 
The corresponding inner product for two velocity fields 
$\boldsymbol{v}, \boldsymbol{w} \in {\cal L}^2 ( \Omega_{ROM} )$ reads
\begin{equation}
\label{eq:innerproduct}
\left( \boldsymbol{v}, \boldsymbol{w} \right)_{ROM} :=
\int\limits_{\Omega_{ROM}} \!\!\!\!  d\boldsymbol{x} \>\> \boldsymbol{v} \cdot \boldsymbol{w}.
\end{equation}
This inner product defines the energy norm
$\Vert \boldsymbol{v} \Vert_{ROM} := \sqrt{\left( \boldsymbol{v}, \boldsymbol{v} \right)}$.

The averaging operation and inner product uniquely 
define the employed snapshot POD \citep{Sirovich1987qam1,Holmes2012book}.
First, the velocity field is decomposed in a mean field, 
$\boldsymbol{u}_0 = \langle \boldsymbol{u} \rangle$, 
and a fluctuating contribution, $\boldsymbol{u}'$, 
following the Reynolds decomposition. 
Then, the  fluctuating part is approximated
by a Galerkin expansion with space-dependent modes $\boldsymbol{u}_i$, $i=1,2,\ldots$ 
and the corresponding mode coefficients $a_i(t)$:
\begin{subeqnarray}
 \label{eq:pod_decomp}
 \boldsymbol{u(\boldsymbol{x}},t)  &=& \boldsymbol{u}_0(\boldsymbol{x}) + \boldsymbol{u}'(\boldsymbol{x},t), \\[3pt]
 \boldsymbol{u}'(\boldsymbol{x},t) &=& \sum_{i=1}^{\infty} a_i(t) \boldsymbol{u}_i(\boldsymbol{x}) \approx  \sum_{i=1}^{N} a_i(t) \boldsymbol{u}_i(\boldsymbol{x}) + \boldsymbol{u}_{res}(\boldsymbol{x},t).
\end{subeqnarray}
POD yields the minimal average squared residual $\left\langle \Vert \boldsymbol{u}_{res} \Vert^2\right \rangle$
as compared to any other Galerkin expansions with $N$ modes \citep{Lumley1970book}.
Note that the snapshot POD method limits the number of POD to $N \le M-1$.
When summing up over $i=1,2,\ldots$, without bound,
we consider the original formulation of POD 
with an accountable infinity of modes.

We re-write the POD expansion more compactly, 
following the convention of \citet{Rempfer1994jfm,Rempfer1994jfm2}:
\begin{equation}
 \label{eq:pod_gsform}
 \boldsymbol{u}(\boldsymbol{x},t) = \boldsymbol{u}_0(\boldsymbol{x}) + \sum_{i=1}^{N} a_i(t) \boldsymbol{u}_i(\boldsymbol{x}) = \sum_{i=0}^{N} a_i(t) \boldsymbol{u}_i(\boldsymbol{x}),
\end{equation}
where $a_0 \equiv 1$,
For later reference, we recapitulate the first and second moments of the POD mode coefficients: 
\begin{equation}
\label{eq:pod_statistics}
\langle a_i \rangle = 0, \quad \langle a_i a_j \rangle = \lambda_i \delta_{ij}.
\end{equation}

The energy content in each mode is given by $K_i(t) = \frac{1}{2}a_i(t)^2$. 
The total turbulence kinetic energy (TKE) resolved by the Galerkin expansion $K_{\Sigma}(t)$ reads 
\begin{equation}
 \label{eq:totalk}
 K_{\Sigma}(t) =\sum_{i=1}^{N} K_i(t).
\end{equation}
The limit $\lim_{N \to \infty} K_{\Sigma}$ for POD yields the TKE $K$ of the velocity field. 
From here and onwards in the paper, the time-averaged value of the quantity $K$, $K_i$ and $K_{\Sigma}$ is implied 
when the $t$ dependence is dropped,
e.g.\ $K = \langle K(t) \rangle$, $K_i = \langle K_i(t) \rangle$, and $K_{\Sigma} = \langle K_{\Sigma}(t) \rangle$. 
Note that by \eqref{eq:pod_statistics}, 
the modal energy and POD eigenvalues are synonymous: $K_i = \lambda_i/2$.

The Galerkin expansion
 \eqref{eq:pod_gsform} satisfies the incompressibility condition by construction. 
The evolution equation for the mode coefficients $a_i$
is derived by a \textit{Galerkin projection} onto the Navier-Stokes equation \eqref{eq:gov_equationsPOD},
i.e.\ from $\left ( \boldsymbol{u}_i, \boldsymbol{R} (\boldsymbol{u}) \right)_{\Omega_{ROM}} = 0$.
Details are provided in the textbooks of \citet{Noack2011book,Holmes2012book}.
For large domains and three-dimensional fluctuations,
the pressure term can generally be neglected
as in \citet{Deane1991pfa}, \citet{Ma2002jfm} and \citet{Noack2005jfm}.
Here, 
the Galerkin projection of the pressure term was found to be negligible 
and it is thus omitted from the model. 
Thus, Galerkin system describing the temporal evolution of the modal coefficients, $a_i(t)$, reads
\begin{equation}
 \label{eq:galerkinsystem}
 \frac{da_i}{dt} = \nu \sum_{j=0}^{N} l_{ij}^{\nu} a_j + \sum_{j,k = 0}^{N} q_{ijk}^c a_j a_k.
\end{equation}
The coefficients $l_{ij}^{\nu}$ and $q_{ijk}^c$ are the \textit{Galerkin system coefficients} 
describing the viscous and convective Navier-Stokes terms, respectively.

\subsection{Hierarchy of low-dimensional Galerkin systems}
\label{Sec:ROM:GS}
In this section,
a subscale-turbulence representations for truncated Galerkin systems are revisited.
We have to account for the dynamic effect of $\boldsymbol{u}_{res}$ of \eqref{eq:pod_decomp}.
First, the exact form of the Galerkin system (propagator) residual is detailed (\S~\ref{Sec:ROM:GS:Residual}).
Then, four eddy viscosity terms for this residual are outlined
based on a single constant eddy viscosity  (\S~\ref{Sec:ROM:GS:GSA}),
a modal constant eddy viscosity (\S~\ref{Sec:ROM:GS:GSB}),
 a single nonlinear eddy viscosity (\S~\ref{Sec:ROM:GS:GSC}),
and a combination of the last two models (\S~\ref{Sec:ROM:GS:GSD}).

\subsubsection{Exact representation of the propagator residual}
\label{Sec:ROM:GS:Residual}
The dynamical system \eqref{eq:galerkinsystem} 
predicts the evolution of all modal coefficients.
By integrating the Galerkin system in time 
we can obtain further long-term information 
about the dynamical behaviour of the original system. 
However, the aim is to simulate the dynamical behaviour of the `large' scales
that presumably govern the global physics of the flow in the wake of the Ahmed body. 
This is desirable since the computational time of computing the convective term, $q_{ijk}^c$, 
and the integration time of the system time scales as $\sim N^3$, 
so that the computational effort soon exceeds the computational effort of the original LES simulation. 
Thus, we want to build a ROM that contains the important physics, 
but with a computational effort to build and 
to integrate in time that is much less than the time of performing the original LES simulations. 
We therefore choose a small number of modes $N$
accounting for the unresolved POD modes at $i=N+1, N+2, \ldots$ 
with a subscale turbulence representation.
Let $\boldsymbol{a} = (a_1, a_2, \ldots, )$ represent the mode coefficients. 
Then the accurate dynamical system takes the following form:
\begin{subeqnarray}
 \label{eq:galerkinsystem2}
 \frac{da_i}{dt} &=& f_i(\boldsymbol{a}) + g_i( \boldsymbol{a}), \\[3pt]
 f_i(\boldsymbol{a}) &=&  
     \nu \sum_{j=0}^N l_{ij}^{\nu} a_j 
       + \sum_{j,k = 0}^N q_{ijk}^c a_j a_k, \\[3pt]
 g_i(\boldsymbol{a}) &=& 
     \nu \sum_{j=N+1}^\infty l_{ij}^{\nu} a_j 
       + \sum_{\substack{j,k = 0 \\ max \{ j,k \} > N  }} ^{\infty} q_{ijk}^c a_j a_k.
 \end{subeqnarray}
Here, the propagator $f_i$ represents the resolved part of the dynamics
while $g_i$ represents the residual of the truncated Galerkin system.
This residual 
contains the viscous and convective terms with at minimum one unresolved mode $i>N$.

In the Kolmogorov description of the turbulence cascade \citep{Kolmogorov1941c,Kolmogorov1941,pope:00},
the large, energy-carrying scales transfers energy to successively smaller scales 
where most of the dissipation of the kinetic energy to internal energy (heat) of the molecules takes place. 
Therefore, any attempt to solve the reduced system in \eqref{eq:galerkinsystem} 
not accounting for the residual, $g_i(\boldsymbol{a})$, 
will lead to excessive energy levels or even divergence of the system. 

\subsubsection{Single constant eddy viscosity (Galerkin system A)}
\label{Sec:ROM:GS:GSA}
In the ground-breaking work by \citet{Aubry1988jfm} on the dynamics of coherent structures in the turbulent boundary layer, 
the residual was modelled by a constant `eddy viscosity' term, 
resulting in a linear subscale turbulence representation $g_i(\boldsymbol{a}) = \nu_0^T \sum_{j=1}^{N} l_{ij}^{\nu} a_j $.
$\nu^T_0$ is generally obtained by solution matching techniques.
In this study, the eddy viscosity is derived from the TKE power balance.
The resulting model \eqref{eq:galerkinsystem2} will be called \textit{Galerkin system A} and abbreviated as GS-A.
 
\subsubsection{Modal constant eddy viscosity (Galerkin system B)}
\label{Sec:ROM:GS:GSB}
\citet{Rempfer1994jfm} refined the linear model by reasoning  that the eddy viscosity
should be scale-dependent resulting in \textit{modal eddy viscosities} $\nu_i^T$, $i=1,\ldots,N$.
The resulting linear subscale turbulence representation reads
$g_i(\boldsymbol{a}) = \nu_i^T \sum_{j=1}^{N} l_{ij}^{\nu} a_j$.
 $\nu_i^T$ can be obtained by solution matching.
In this study, $\nu_i^T$ is derived from the modal power balance \citep{Noack2005jfm}. 
We refer to the resulting model as \textit{Galerkin system B}, or GS-B.

\subsubsection{Single nonlinear eddy viscosity (Galerkin system C)}
\label{Sec:ROM:GS:GSC}
\citet{Noack2011book} remark that the subscale turbulence representations of GS-A and GS-B are linear
while the energy transfer is caused by nonlinear mechanisms.
We start with a single eddy viscosity ansatz
\begin{equation}
 \label{eq:modalansatz}
 g_i(\boldsymbol{a}) = \nu^T_0 (\boldsymbol{a}) \sum\limits_{j=1}^N l_{ij}^{\nu} a_j,
\end{equation}
but allow the eddy viscosity to be state dependent.
On the other hand
\eqref{eq:galerkinsystem2} is written as
\begin{equation}
 \label{eq:gsresidualansatz}
 g_i(\boldsymbol{a}) = 
  \nu \sum_{j=N+1}^{\infty} l_{ij}^{\nu} a_j 
+ \sum_{\substack{j,k = 0 \\ max \{ j,k \} > N}}^{\infty} q_{ijk}^c a_j a_k.
\end{equation}
Evidently, both terms cannot be exactly matched.
However,  the energy transfer rate effect should be similar.
In the  modal power balance, this energy loss is quantified with $\langle a_i g_i \rangle$.
Equality of the energy transfer rate yields
\begin{equation}
\label{eq:gsresidualansatz2}
  \nu^T l_{ii}^{\nu} \lambda_i
= \sum_{\substack{j,k = 0 \\ max \{ j,k \} > N } }^{\infty} T_{ijk}, \text{ where } T_{ijk} =  q_{ijk}^c \langle a_i a_j a_k \rangle,
\end{equation}
exploiting $\langle a_i a_j \rangle = \lambda_i \delta_{ij}$ \eqref{eq:pod_statistics}.
The triadic power terms on the right-hand side may be approximated with a finite-time thermodynamics closure
\citep{Noack2008jnet}
\begin{equation}
 \label{eq:tijkftt}
 T_{ijk} = \alpha \chi_{ijk} \sqrt{K_i K_j K_k}\left(1 - \frac{3K_i}{K_i + K_j + K_k} \right)
.
\end{equation}
In the next step,  we introduce relative modal energy contents $\kappa_i$ via
$K_i = \kappa_i K_{\Sigma}$, with $\sum_{i=1}^N \kappa_i$ = 1. 
Then, \eqref{eq:gsresidualansatz2} becomes
\begin{equation}
 \label{eq:gsresidualansatz3}
 2\nu^T l_{ii}^{\nu} \kappa_i = 
 \sqrt{K_{\Sigma}} \> \sum_{\substack{j,k = 0 \\ max \{ j,k \} > N } }^{\infty}
 \alpha  \> \chi_{ijk}  \sqrt{\kappa_i \kappa_j \kappa_k} 
                 \left(1 - \frac{3 \kappa_i}{ \kappa_i + \kappa_j + \kappa_k } \right).
\end{equation}
This closure relation suggests that $\nu_T$ scales with $\sqrt{K_{\Sigma}}$,
assuming that $\kappa_i$ remain approximately constant with $K_{\Sigma}$.
The resulting nonlinear eddy viscosity model used in the present work 
thus takes the form:
\begin{equation}
 \label{eq:nonlinearmodel}
 g_i(\boldsymbol{a}) = \nu_0^T \sqrt{ \frac{K_{\Sigma}(t)}{K_{\Sigma}} } \sum_{j=1}^{N} l_{ij}^{\nu} a_j.
\end{equation}
Thus, large (small) fluctuation levels $K_{\Sigma}(t) > K_{\Sigma}$  ($K_{\Sigma}(t) < K_{\Sigma}$) 
lead to a higher (smaller) damping than predicted by the corresponding linear subscale turbulence representation.
In particular, boundedness of the new \textit{Galerkin system C} (GS-C) can be proven,
if the energy preservation of the quadratic term is enforced \citep{Cordier2013ef}.

\subsubsection{Modal nonlinear eddy viscosity (Galerkin system D)}
\label{Sec:ROM:GS:GSD}
Combining the nonlinear eddy viscosity of GS-C \eqref{eq:nonlinearmodel}
and the modal eddy viscosities of GS-B 
yields the following nonlinear subscale turbulence representation:
\begin{equation}
 \label{eq:modalnonlinearmodel}
 g_i(\boldsymbol{a}) = \nu_i^T \sqrt{ \frac{K_{\Sigma}(t)}{K_{\Sigma}} } \sum_{j=1}^{N} l_{ij}^{\nu} a_j.
\end{equation}
The resulting dynamical system 
is referred to as \textit{Galerkin system D}, or GS-D.

\section{Results}
\label{Sec:Results}
In this section, 
we present results of the four Galerkin systems A--D
from \S~\ref{Sec:ROM:GS:GSA}, \S~\ref{Sec:ROM:GS:GSB}, \S~\ref{Sec:ROM:GS:GSC} and \S~\ref{Sec:ROM:GS:GSD}, respectively.
First (\S~\ref{Sec:Results:POD}), the POD is presented.
The solutions of Galerkin systems A--D 
are compared (\S~\ref{Sec:Results:GSX}).
Finally (\S~\ref{Sec:Results:Robustness}), 
the robustness of the Galerkin systems in terms of model parameters is investigated.

\subsection{POD}
\label{Sec:Results:POD}

\begin{figure}
  \centerline{(\textit{a})\includegraphics[height=50mm,trim=0cm 0cm 0cm 0cm,clip=true,keepaspectratio]{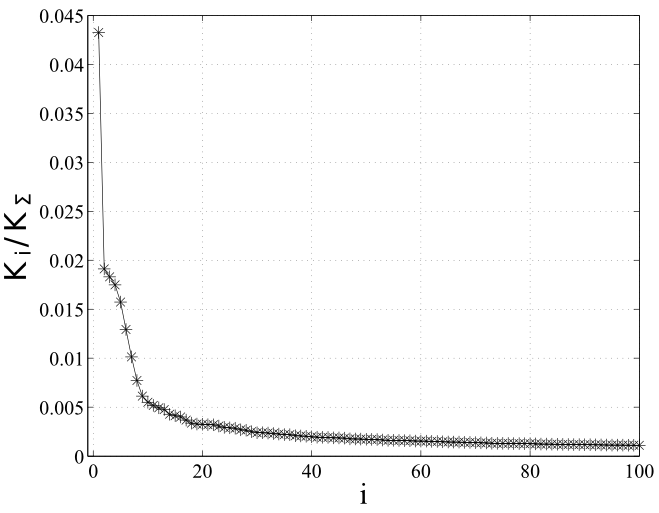}(\textit{b})\includegraphics[height=50mm,trim=0cm 0cm 0cm 0cm,clip=true,keepaspectratio]{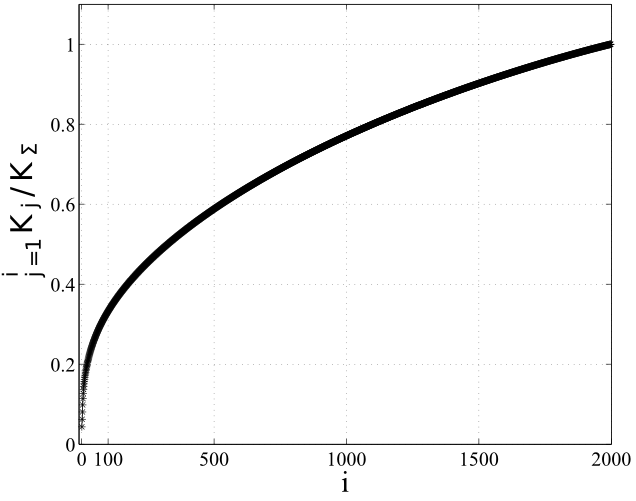}}
  \caption{Spectrum from the POD:
(\textit{a}) Normalized spectrum; (\textit{b}) Normalized cumulative spectrum.
The first mode has by far the largest energy level.
It resolves asymmetric base-flow variations between positive and negative side forces.}
\label{fig:4}
\end{figure}

Figure \ref{fig:4} presents the cumulative spectrum 
from the POD eigenvalues of the dataset of flow snapshots during the considered time-interval. 
The convergence rate is quite slow, and the first 100 modes contain some 35$\%$ of the total kinetic energy in the system.
The first 500 modes resolve 60$\%$ of the kinetic energy.
We shall not pause to visualize the POD modes,
as the structure of the modes contributes little 
to the understanding of the subscale turbulence representations.
It is only important to note that the first POD mode $\boldsymbol{u}_1$
describes a slow base-flow change between positive and negative side forces.
This mode shall be called \textit{shift mode} \citep{Noack2003jfm}
reflecting the analogous role in resolving base flow changes.

\subsection{Comparative study of the Galerkin systems}
\label{Sec:Results:GSX}

\begin{figure}
\centerline{\includegraphics[height=50mm,trim=0cm 0cm 0cm 0cm,clip=true,keepaspectratio]{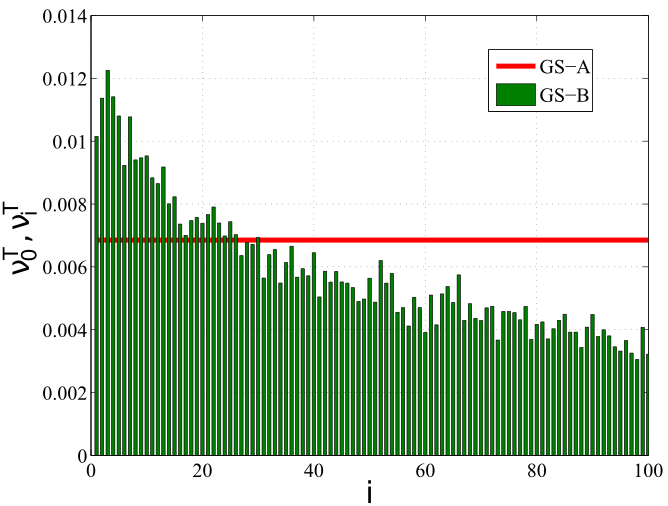}}
\caption{Total and modal eddy viscosities of Galerkin systems A and B.}
\label{fig:5}
\end{figure}

The key parameters of the subscale turbulence representation
are the total and modal eddy viscosities of Galerkin systems A and B, respectively.
These parameters have been determined by the total and modal power balance
for GS-A and GS-B, respectively.
In other words, no solution matching is performed.
Figure \ref{fig:5} shows their values.
The total eddy viscosity $\nu_0^T$ lies between the extremal modal values, as expected.
The modal eddy viscosities $\nu_i^T$ are all positive and 
follow a nearly monotonous trend with the mode index $i$.
Such a nearly monotonous behaviour indicates a good quality of the LES data.
For other flow data, the authors frequently observe a large scatter of these values with $i$.

Galerkin system C (GS-C) assumes $\nu_0^T$ of GS-A 
and rescales the value according to the square-root law \eqref{eq:nonlinearmodel}.
Similarly, GS-D applies the same scaling to the modal eddy viscosities $\nu_i^T$ of GS-B.

\begin{figure}
\centerline{\includegraphics[height=50mm,trim=0cm 0cm 0cm 0cm,clip=true,keepaspectratio]{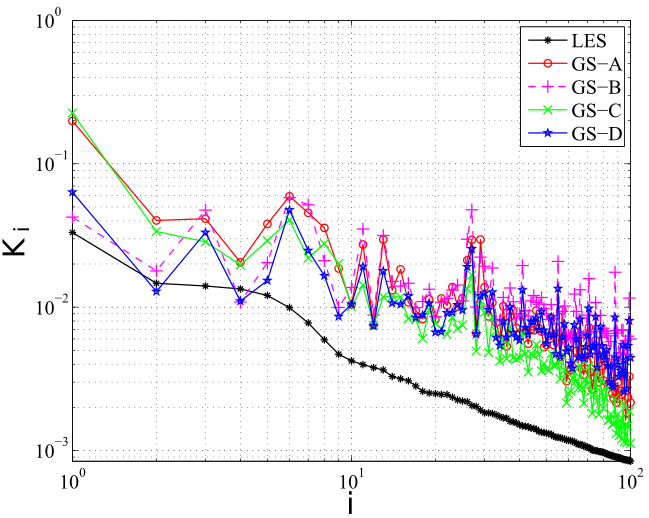}}
\caption{Comparison of the modal kinetic energies  $K_i$ between the LES simulation 
and the four Galerkin system solutions (GS-A$\ldots$D).}
\label{fig:6}
\end{figure}
We start the comparison of the four Galerkin systems
with the modal energy spectrum $K_i$ of their respective long-term solutions.
From figure \ref{fig:6}, 
the most simple GS-A is seen to deviate strongest from the CFD values.
Better spectra might be obtained with solution matching techniques for $\nu_0^T$ (see below),
but such a procedure indicates that the total power balance as consistency condition is violated.

GS-B indicates an increased performance 
by replacing the total eddy viscosity by the modal analogues.
The increased employed knowledge from the Navier-Stokes equation,
namely the use of $N$ modal power balances, pays of.

GS-C tends to outperform both Galerkin systems,
particularly for large mode indices.
This indicates that the nonlinearity of the eddy viscosity ansatz 
is a crucial physical enabler and should not be ignored.
The correct TKE-dependent scaling of the eddy viscosity 
appears to be more important 
than the modal refinement of their values.
However, one should note 
that the deviation of the modal eddy viscosity values
from the total analogue is less than a factor 2
for this particular flow.
The energy levels in mode 1 (the shift mode) 
and the oscillatory modes $i=2,\ldots,11$
are over-predicted 
by GS-C as compared to the levels of the LES. 
One reason for this over-prediction is a short-coming of the total eddy viscosity:
The physically correct modal values for these modes are almost 2 times larger. 
Similarly, the next (large-scale) oscillatory POD modes $i=2,\ldots, 7$
are over-predicted.

Galerkin system D has larger eddy viscosities 
for the most dominant first POD modes and 
cures the over-prediction of the corresponding modal amplitudes of GS-C.
The modal energies of the higher-order modes of GS-B and GS-D are comparable.\textcolor{black}{
The modal refinement of the nonlinear eddy viscosity term has a price:
the higher-order (less energetic) modes of GS-D 
tend to be more energetic than the ones of GS-C.
We hypothesize that the cause is 
a broad-band frequency time-variation of the eddy viscosity due to $K_{\sum}(t)$.
The modal eddy viscosities of the higher-order modes are smaller than 
the total eddy viscosity (see figure \ref{fig:5}).
Hence, the higher-order modes of GS-D are less damped and more energized 
by the unsteady subscale turbulence term
as compared to GS-C.
A low-pass filter on $K_{\sum}(t)$ could cure this problem,
if the low-energy tail of the POD decomposition is of sufficient interest.
We shall not pause to incorporate this additional refinement.}

\begin{figure}
\centerline{(\textit{a})\includegraphics[height=50mm,trim=0cm 0cm 0cm 0cm,clip=true,keepaspectratio]{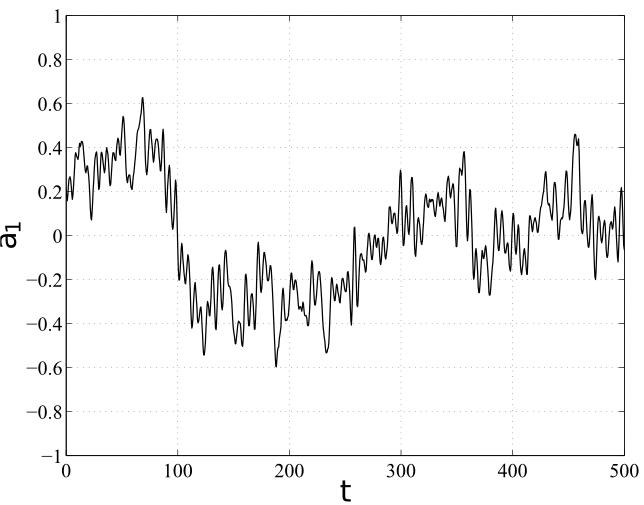} (\textit{b})\includegraphics[height=50mm,trim=0cm 0cm 0cm 0cm,clip=true,keepaspectratio]{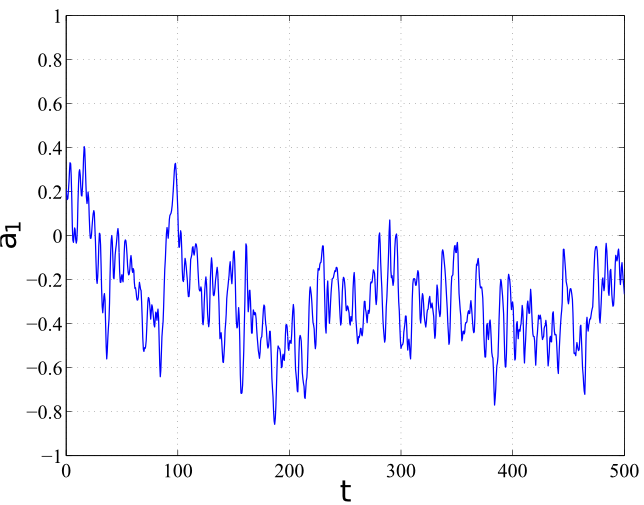}}

\centerline{(\textit{c})\includegraphics[height=50mm,trim=0cm 0cm 0cm 0cm,clip=true,keepaspectratio]{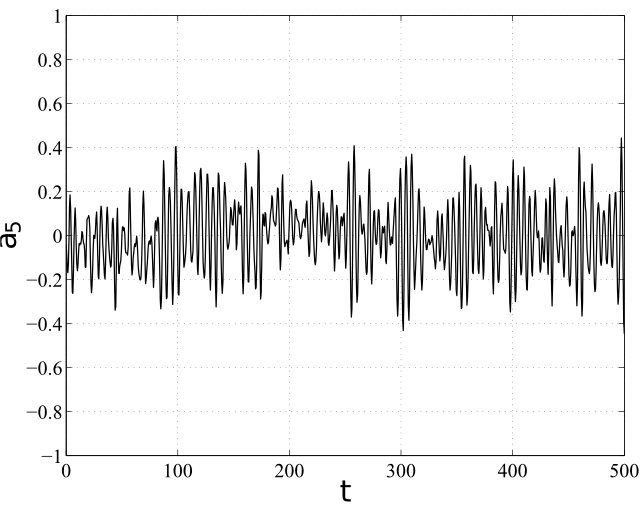} (\textit{d})\includegraphics[height=50mm,trim=0cm 0cm 0cm 0cm,clip=true,keepaspectratio]{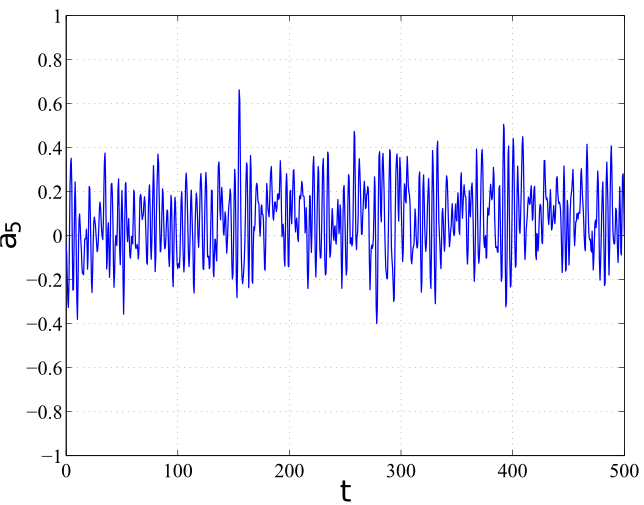} }


\centerline{(\textit{e})\includegraphics[height=50mm,trim=0cm 0cm 0cm 0cm,clip=true,keepaspectratio]{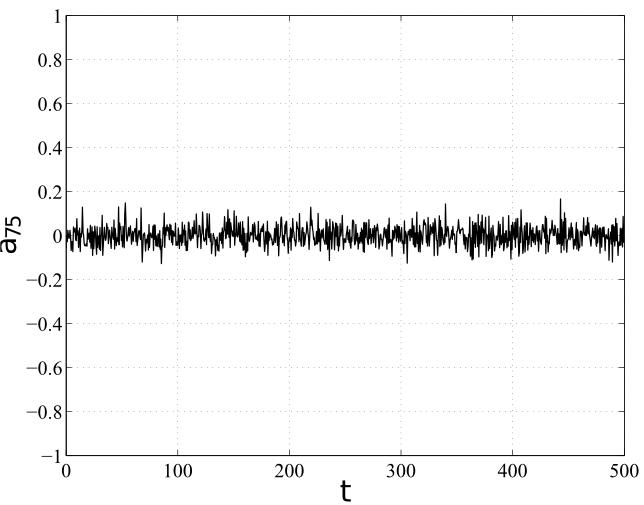} (\textit{f})\includegraphics[height=50mm,trim=0cm 0cm 0cm 0cm,clip=true,keepaspectratio]{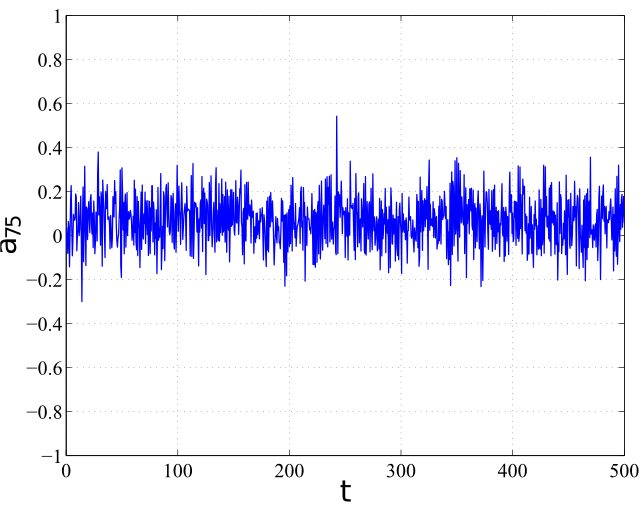} }

\caption{Evolution of the mode coefficients $a_i(t)$.
Comparison between LES simulation (left)  and GS-D (right) for $N = 100$. 
a,(\textit{b}) $a_1(t)$ describing asymmetric base flow changes (shift mode);
c,(\textit{d}) $a_5(t)$ as example of a dominant oscillatory POD mode; 
e,(\textit{f}) $a_{75}(t)$ representing a higher-order POD mode.}
\label{fig:7}
\end{figure}
The temporal dynamics of the GS-D and  the POD (from LES data)
are presented for selected modes in figure \ref{fig:7}. 
The first coefficient $a_1$ of the shift mode is the most interesting one.
This coefficient is depicted in figure \ref{fig:7}(\textit{a}) 
and describes the change from  one asymmetric base-flow state to the other. 
Its value follows exactly
the side-force signal from the LES simulation in figure \ref{fig:2}(\textit{b}).
Only  Galerkin system D was found capable of predicting the sudden switches 
from one side force state to the other with realistic amplitudes at realistic time scales.
Also GS-C exhibits such base-flow changes,
but the amplitude is over-predicted by a factor 2
and these time scales were over-predicted by three orders of magnitudes.
GS-A predicts a purely periodic solution for $N = 100$, and 
GS-B does not predict the amplitude in a physical correct way.
Summarizing, 
both, the modal refinement of the eddy-viscosity (GS-B and GS-D) 
and their energy-dependent scaling in (GS-C and GS-D) 
emerge as crucial enablers for the accurate Galerkin systems.  

\subsection{Robustness study of the Galerkin systems}
\label{Sec:Results:Robustness}
In this section, 
the robustness of the Galerkin systems
with respect to their dimension and 
the eddy viscosity parameters is investigated.

\begin{figure}
\centerline{\includegraphics[height=50mm,trim=0cm 0cm 0cm 0cm,clip=true,keepaspectratio]{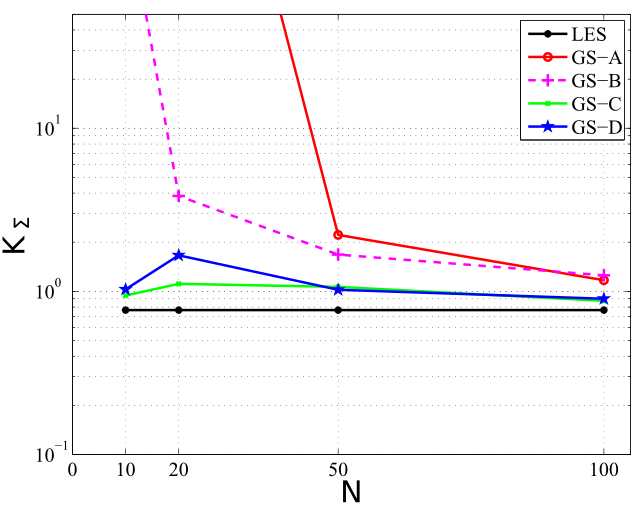}}
\caption{Comparison of the total energy, $K_{\Sigma}$, 
in the ROM for different dimensions $N$ of the ROM. The eddy viscosity is kept constant to the value of that of $N=100$.}
\label{fig:8}
\end{figure}
In figure \ref{fig:8},
the time-averaged total energy of all Galerkin systems
is depicted  for different dimensions $N$ of the ROM.
GS-D has, on average, the best agreement with the LES values
for all 4 dimensions, i.e.\ $N=10, 20, 50$ and $100$.
In contrast, the most simple GS-A shows even the wrong trend with respect to $N$. 
We emphasize that all eddy viscosity values are derived from TKE power balances.
The performance of each Galerkin system could easily be improved
with solution matching techniques for these parameters.
However, the price of such techniques 
is a potentially large residual in the TKE power balance,
i.e.\ the predicted modal energy distribution and energy flows may be significantly distorted.

\begin{figure}
\centerline{\includegraphics[height=50mm,trim=0cm 0cm 0cm 0cm,clip=true,keepaspectratio]{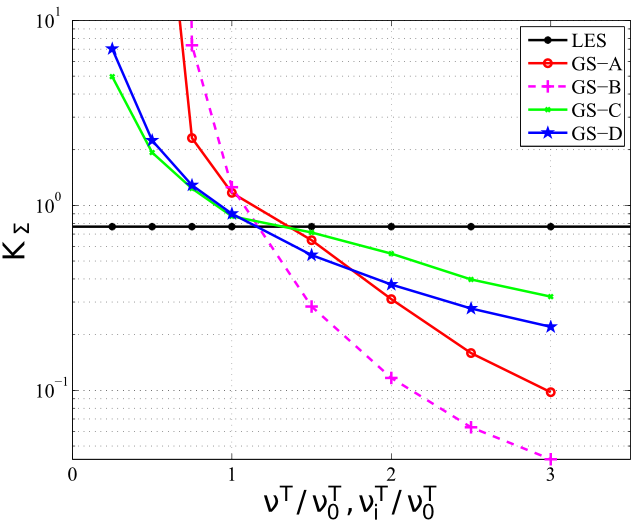}}
\caption{Comparison of the total energy, $K_{\Sigma}$, in the ROM for different values of $\nu^T_{i}$ and $\nu^T_{0}$.}
\label{fig:9}
\end{figure}
Finally, 
the role of the eddy viscosity parameter 
is investigated in figure \ref{fig:9} for $N=100$. 
For the simulations in this figure, 
we have varied the total viscosities  $\nu^T_i$
in the range 25$\,\%$, 50$\,\%$, 75$\,\%$, 150$\,\%$, 200$\,\%$, 250$\,\%$ and 300$\,\%$ 
of the reference value $\nu_0^T$ for GS-A and GS-C.
The modal eddy viscosities $\nu_i^T$ of GS-B and GS-D  
have been changed by the same factor depicted on the abscissa.
GS-A does not show monotonous behaviour in terms of the parameter change.
GS-B has a more physical monotonous behaviour but its deviations are stronger than for GS-A.
One may speculate that modal eddy viscosities are 'over-fitted' for the physical reference level.
GS-C is seen to be far less affected by these changes,
the nonlinear eddy viscosity compensates for too small or too large reference values.
Finally, GS-D follows GS-C but shows an even smoother curve,
thus indicating the largest robustness.

\begin{figure}
\centerline{(\textit{a})\includegraphics[height=50mm,trim=0cm 0cm 0cm 0cm,clip=true,keepaspectratio]{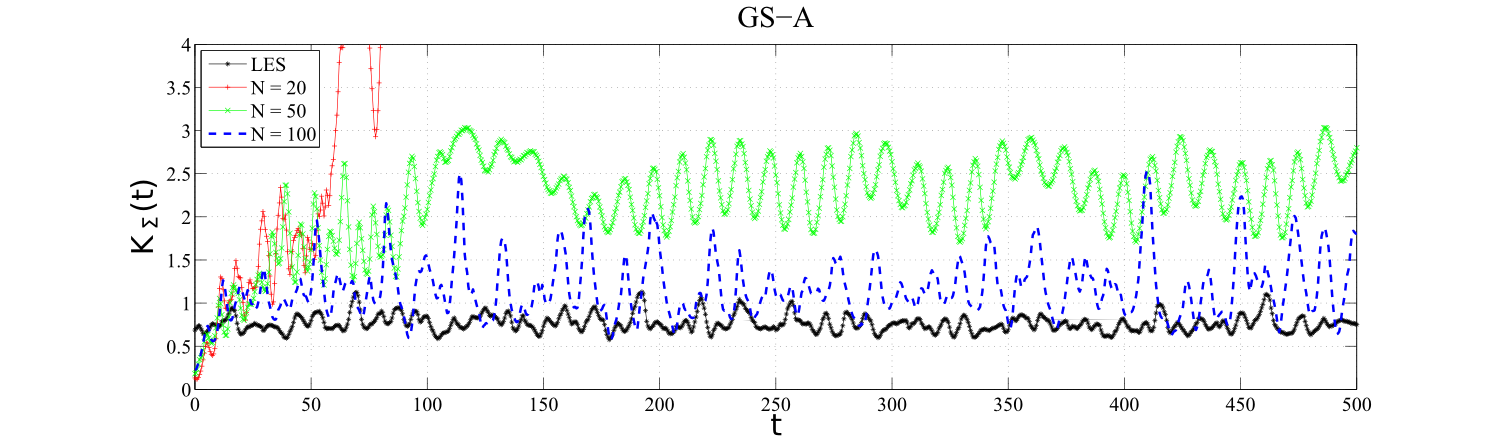}}
\centerline{(\textit{b})\includegraphics[height=50mm,trim=0cm 0cm 0cm 0cm,clip=true,keepaspectratio]{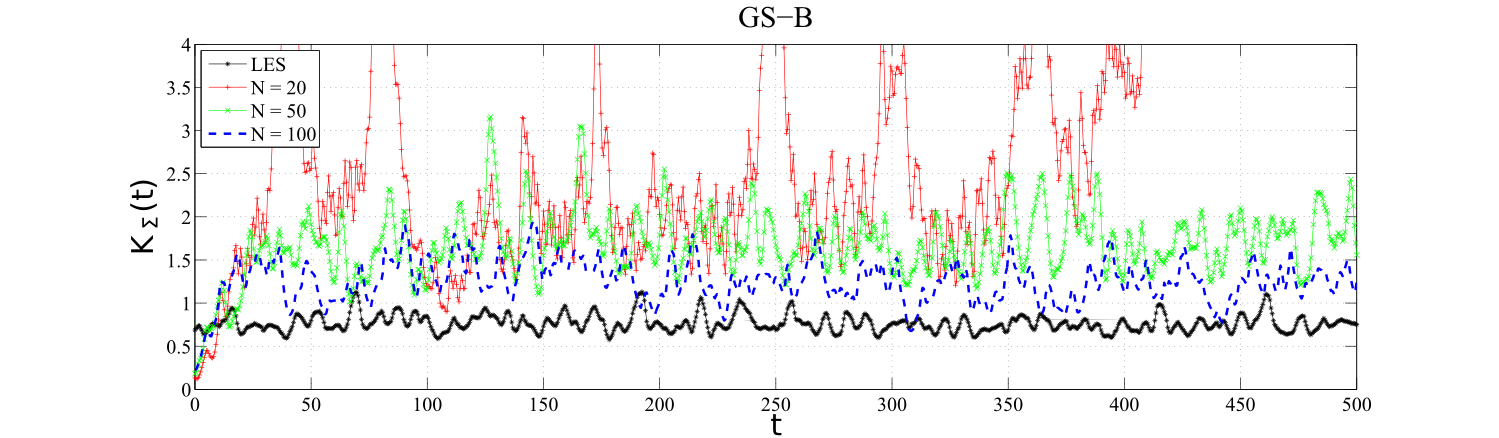}}
\centerline{(\textit{c})\includegraphics[height=50mm,trim=0cm 0cm 0cm 0cm,clip=true,keepaspectratio]{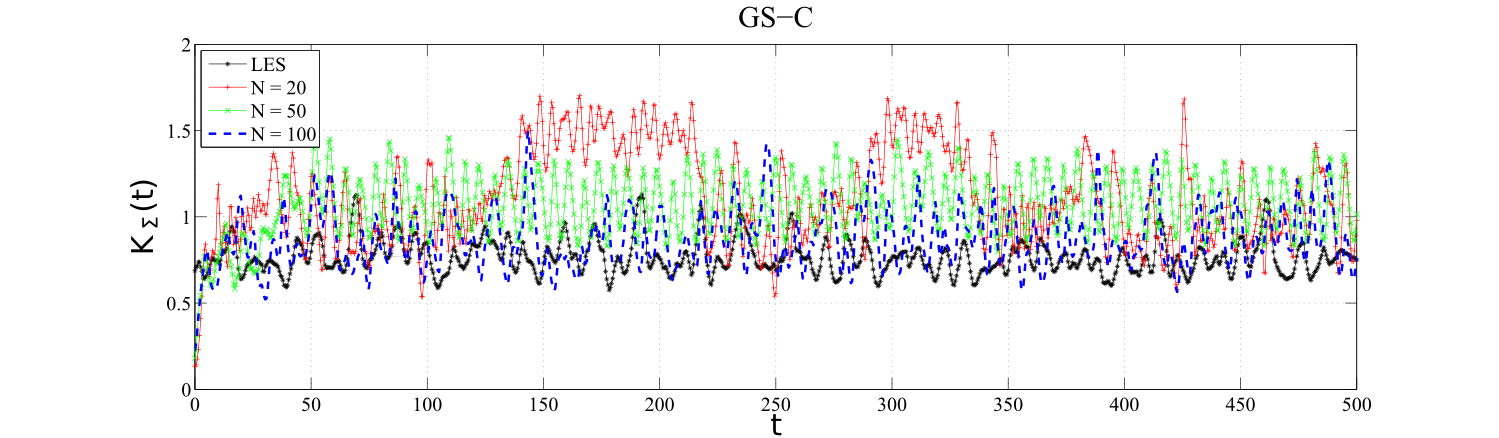}}
\centerline{(\textit{d})\includegraphics[height=50mm,trim=0cm 0cm 0cm 0cm,clip=true,keepaspectratio]{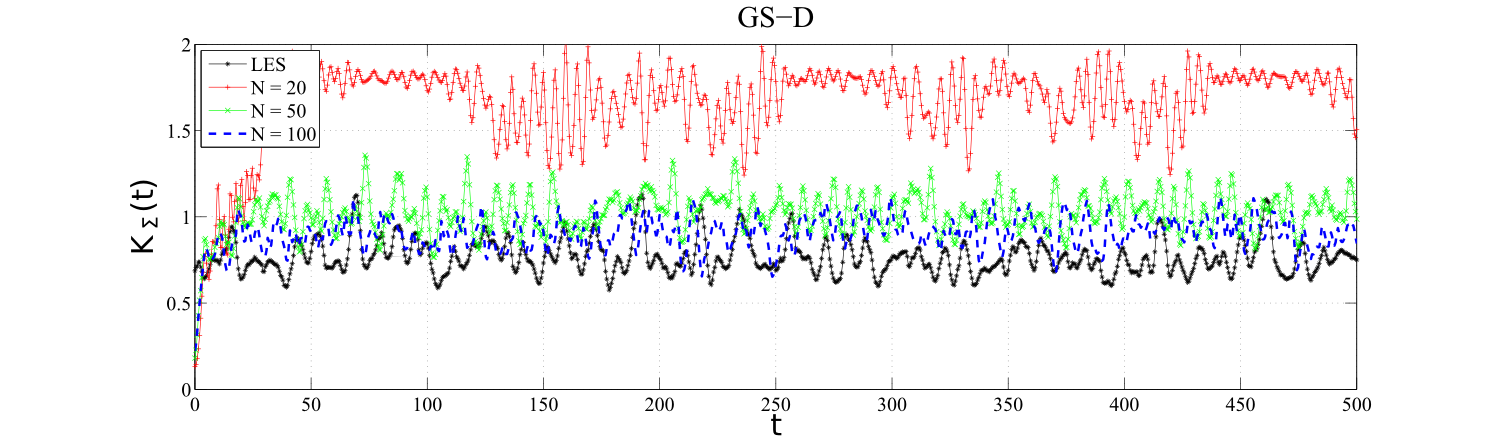}}
\caption{\textcolor{black}{Comparison of instantaneous total energy $K_{\Sigma}(t) = \sum_{i=1}^{N} K_i(t)$ 
of the Galerkin systems:
(\textit{a}) GS-A; 
(\textit{b}) GS-B; 
(\textit{c}) GS-C; 
(\textit{d}) GS-D. Note that the scaling of the $y$-axis between figures (\textit{b}) and (\textit{c}) is different.}}
\label{fig:10}
\end{figure}
Finally, in figure \ref{fig:10} we present the temporal evolution 
of the total energy level for all four Galerkin systems
for  dimensions $N=20$, $N=50$, $N=100$. 
The systems with dimensions $N=20$ and $N=50$
are pure truncations of the $N=100$ reference.
This implies that the eddy viscosities of the $N=100$ reference are kept constant 
in this system reduction.
We do not want to mix the effect of varying dimensions \textit{and} varying eddy viscosity. 
Again, GS-C and GS-D with nonlinear subscale turbulence representation 
outperform GS-A and GS-B in terms of robustness.

\textcolor{black}{Mean values and the variances of the signals from the $N=100$ ROMs (see figure \ref{fig:10}) and LES are presented in table \ref{tab:stats}. GS-C predicts the mean value slightly closer to LES than GS-D, but the variance of GS-C (0.0304) is overpredicted by a factor of three to that compared to the LES (0.0095), while the variance of GS-D (0.0091) is close to that of the LES. GS-A and GS-B overpredict the mean and variance significantly.}

\begin{table}
  \textcolor{black}{
  \begin{center}
\def~{\hphantom{0}}
  \begin{tabular}{lccccc}
   & GS-A & GS-B & GS-C & GS-D & LES \\ 
Mean & 1.1693 & 1.2535 &  0.8740 & 0.9010 & 0.7671 \\
Variance & 0.1264 & 0.0606 & 0.0304 & 0.0091 & 0.0095 \\
  \end{tabular}
  \caption{Mean values and variances of $K_{\sum}(t)$ for the N = 100 ROMs and LES.}
  \label{tab:stats}
  \end{center}
  }
\end{table}

In summary, the accuracy and robustness of the Galerkin system 
is found to improve by modal refinement of the eddy viscosities
and by an energy-dependent scaling.
A similar observation for the energy-dependent scaling
has been made for the POD models of a mixing layer by \citet{Cordier2013ef}.

\section{Conclusions}
\label{Sec:Conclusions}
We have investigated a hierarchy 
of linear and nonlinear eddy viscosity terms 
for a POD Galerkin model
accounting for the unresolved velocity fluctuations.
The chosen configuration is a high-Reynolds number flow 
over a square-back Ahmed body.
This flow exhibits
three challenging features for reduced-order models.
Firstly, 
the high Reynolds number implies 
that a subscale turbulence representation
is mandatory for realistic fluctuation levels,
or even boundedness, of the Galerkin system solution.
Secondly, 
the coherent structures of the Ahmed body have a broadband frequency signature.
The resulting frequency cross-talk implies 
that many modal interactions exist 
and need to be correctly resolved.
And thirdly,
the base flow has two meta-stable states 
with nearly constant non-vanishing side forces.
Experiments of a similar Ahmed body configuration \citep{Grandemange2013jfm}
exhibit these asymmetric quasi-attractors.
Such quasi-attractors imply a complex interaction from small to very large time scales
and constitute a significant modelling challenge 
--- even for Navier-Stokes simulations.

The solutions of a POD Galerkin model with 100 modes or less 
converge to infinity 
underlining the need for a subscale turbulence presentation.
Four corresponding auxiliary models have been tested,
using a single or modally refined eddy viscosities
with constant or energy-dependent values.
These parameters are determined from the total or modal TKE power balance.
Solution matching techniques are excluded as parameter identification method,
to avoid any inconsistency with the TKE power balances.
A single constant eddy viscosity, as used by \citet{Aubry1988jfm} and others,
is already sufficient to stabilize the Galerkin solution.
Modally refined viscosities, as suggested by \citet{Rempfer1994jfm2},
are found to significantly improve the accuracy of the modal fluctuation energies.
Yet, both approaches rely on constant eddy viscosities
leading to linear subscale turbulence representations for nonlinear energy flow cascade. 
The resulting Galerkin solutions
converge to infinity if the initial conditions are far from the attractor.
In addition, both Galerkin systems do no exhibit 
the meta-stable asymmetric base flow states.

We corroborate the need of eddy viscosities
which scale with the square-root of the resolved fluctuation energy.
The single nonlinear eddy viscosity model leads to an accurate prediction
for the fluctuation levels of higher-order modes,
while the amplitudes of the first 7 modes are over-predicted.
Arguably, the first 7 modes define the large-scale coherent structures 
and are the most important part of the spectrum.
The modally refined eddy viscosity cures this over-prediction
at the expense of a less accurate tail of the modal energy spectrum.
These nonlinear eddy viscosity models 
are capable of resolving the flipping  between asymmetric base flow states.
In addition, the resulting Galerkin systems convergence to their respective attractors
for initial conditions --- even if these  are far away from them.
Global converge can strictly be ensured by enforcing the energy preservation on the quadratic term.
This energy preservation is derivable from the Navier-Stokes equation \citep{Kraichnan1989physd,Schlegel2013jfm}.
In addition, Galerkin systems with nonlinear subscale turbulence representations
are shown to be much more robust with respect to changes of the eddy viscosity parameters
and the dimension of the model.

The modally refined, nonlinear eddy viscosity terms have significantly increased accuracy
and robustness of the Galerkin system 
as compared to traditional linear subscale turbulence representations.
The accuracy has been achieved with a parameter identification,
 purely based on Navier-Stokes equation based constraints 
and without solution matching techniques.
The robustness is a key enabler for three ROM-based applications.
Firstly, ROM may serve as test-bed for the understanding of the nonlinear dynamics.
One key question is the mechanism for the amplitude selection,
i.e.\ what drives the transients towards the attractor.
Secondly, ROM may be employed 
as a computationally inexpensive surrogate model for multiple purposes, 
e.g.\ for the inlet conditions of the flow around a following car model.
In this case, it is desirable to have ROM 
which works over a certain range of operating conditions,
e.g.\ slowly varying oncoming velocity.
This variability implies that the ROM employs 
a physically correct robust amplitude selection mechanism,
e.g.\ does not diverge for a small change of the Reynolds number.
Finally, model-based control design requires a ROM 
which works robustly for a range of natural and forced transients.
Moreover, the control design is often based on a hierarchy
of ROM with different dimension
--- ranging from robust least-order models 
to more accurate higher-order models,
which pose larger challenges to state estimation.
The nonlinear eddy viscosity term 
serve all three mentioned applications.
\textcolor{black}{Table \ref{tab:conclusion} summarizes the achieved benefits from the modal and nonlinear eddy viscosity refinements.}
\begin{table}
\textcolor{black}{
  \begin{center}
\def~{\hphantom{0}}
  \begin{tabular}{c@{\quad}l}
 \multicolumn{2}{l}{\bf{Constant eddy viscosity (GS-A)}} \\
          $-$ &blow up in finite time for certain initial conditions \\
          $-$ &significant overprediction of fluctuation levels \\
          &   (with Navier-Stokes equation inferred eddy viscosity) \\
\multicolumn{2}{l}{\bf{Modal eddy viscosities (GS-B)}} \\
          -&blow up in finite time for certain initial conditions \\
          +&more accurate prediction of fluctuation levels \\
          +&reduced dependency on ROM dimension $N$ \\
\multicolumn{2}{l}{\bf{Non-linear constant eddy viscosity (GS-C)}} \\
          -&significant overprediction of fluctuation levels \\
          +&guaranteed boundedness of solution independent of the initial condition  \\
          +&reduced dependency on ROM dimension $N$ \\
          +&reduced dependency on $\nu_T$ variation \\
\multicolumn{2}{l}{\bf{Non-linear modal eddy viscosity (GS-D)}} \\
          +&guaranteed boundedness of solution independent of the initial condition  \\
          +&more accurate prediction of fluctuation levels \\
          +&reduced dependency on ROM dimension (N) \\
          +&reduced dependency on $\nu_T$ variation 
  \end{tabular}
  \caption{Performance of the Galerkin systems A--D: 
           '$-$' refers to challenges, 
           '$+$' to improvements with respect to the benchmark Galerkin system A.}
  \label{tab:conclusion}
  \end{center}
  }
\end{table}

To conclude, the proposed nonlinear subscale turbulence term
with modal eddy viscosity of \citet{Rempfer1994jfm2}
and energy-dependent scaling of \citet{Noack2011book}
is a recipe for accurate and robust POD models 
for a large class of complex flows, 
comprising the flow over an Ahmed body as shown here,
a mixing layer \citep{Cordier2013ef}, 
and subsonic jet noise \citep{Schlegel2009springer}.
The study emphasizes the decisive role 
of a good structure identification of the Galerkin system propagator
--- here in form of nonlinear stabilizing term ---
before parameter identification methods are to be applied.

\section{Acknowledgements}
The thesis of J. \"Osth is supported financially by Trafikverket (Swedish Transport Administration).
The thesis of D. Barros is supported financially by PSA and ANRT in the context of the OpenLab Fluidics between PSA and Institute Pprime.
The work by S. Krajnovi\'c in this paper was partially funded by the Chalmers Sustainable Transport Initiative.
The authors acknowledge the funding and excellent working conditions 
of the Senior Chair of Excellence
'Closed-loop control of turbulent shear flows 
using reduced-order models' (TUCOROM)
supported by the French Agence Nationale de la Recherche (ANR)
and hosted by Institute PPRIME.
We thank the Ambrosys Ltd.\ Society for Complex Systems Management,
the Bernd Noack Cybernetics Foundation 
and OpenLab PPRIME/PSA for additional support.
We appreciate valuable stimulating discussions 
with our collaborators
Markus Abel, 
Jean-Paul Bonnet, 
Steven Brunton,
Laurent Cordier,
Jo\"el Delville, 
Thomas Duriez, 
Fabien Harambat,
Eurika Kaiser,
Robert Niven, 
Tamir Shaqarin,
Vladimir Parezanovic,
Bartek Protas, 
Tony Ruiz,
Michael Schlegel, 
Marc Segond and
Andreas Spohn.
Special thanks are due to Nadia Maamar 
for a wonderful job in hosting the TUCOROM visitors.

Software licenses for AVL Fire were provided by AVL List GmbH. 
Computations were performed at SNIC (Swedish National Infrastructure for Computing) at the Center for Scientific Computing at Chalmers (C3SE), Center for High Performance Computing at KTH (PDC) and National Supercomputer Center (NSC) at LiU.

Last but not least, we thank the referees
for their constructive suggestions.

\appendix
\section{\textcolor{black}{Comparison between experimental data and LES data}}\label{appA}
This appendix describes the companion experiment
at Institute PPRIME, which serves as a reference for the LES simulation. PIV and hot-wire data
are only used to validate the data obtained from the LES simulation.

\subsection{Description of experimental set-up}
\begin{figure}
  \centerline{\includegraphics[width=80mm,trim=0cm 0cm 0cm 0cm,clip=true,keepaspectratio]{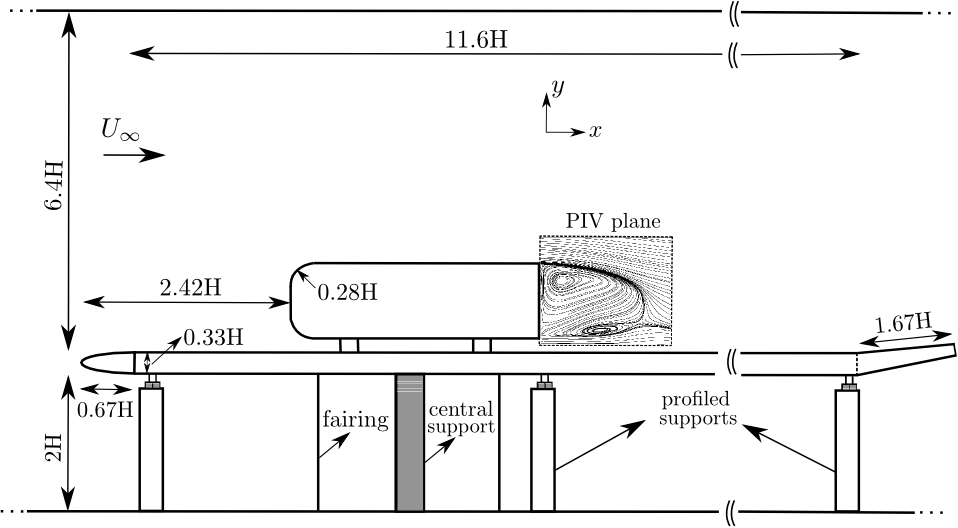}}
  \caption{Experimental set-up.}
\label{fig:expsetup}
\end{figure}
The experiments were conducted in a closed-loop wind tunnel with a test section of 6.24
m$^2$. The model was mounted over an elliptical leading-edge flat plate as illustrated in figure \ref{fig:expsetup}.
At the end of the flat plate, an inclined flap was adjusted in order to obtain an upstream flow
aligned perpendicularly to its leading edge. This procedure was done without the bluff-body
in the wind tunnel. Considering the upper area above the plate, the blockage ratio is about
2$\%$ and no blockage corrections were performed. The upstream velocity, measured on the upper
surface of the wind tunnel (above the model), was kept constant and equals to $U_{\infty}$ = 15$\,$m/s.
Particle image velocimetry (PIV) was performed on the near wake (see detail in figure \ref{fig:expsetup}).
Streamwise and transverse (respectively $x$ and $y$ directions) components of the velocity field
were measured using two LaVision Imager pro X 4M (resolution 2000x2000 pixels) cameras. A
laser sheet was pulsed (with time delays of 120$\,\mu$s) in the symmetry plane of the configuration
and image pairs were acquired at a sampling frequency of $3\,$Hz. Velocity vectors calculations
are processed with an interrogation window of 32x32 pixels (an overlap of 50$\%$) giving a spatial resolution of about 1$\%$ of the model’s height. Starting with an
absolute displacement error of 0.1 pixels, the maximum uncertainty on instantaneous velocity
fields are estimated to be 0.2$\,$m/s. The mean flow was computed using 500 independent
velocity fields and the estimated statistical error for time-averaged velocity is 0.09 $\sigma_{rms}$ with
95$\%$ of confidence level and $\sigma_{rms}$ is the
local root mean square of the velocity.

\subsection{Velocity profiles}
\begin{figure}
\centerline{(\textit{a})\includegraphics[height=50mm,trim=0cm 0cm 0cm 0cm, clip=true, keepaspectratio]{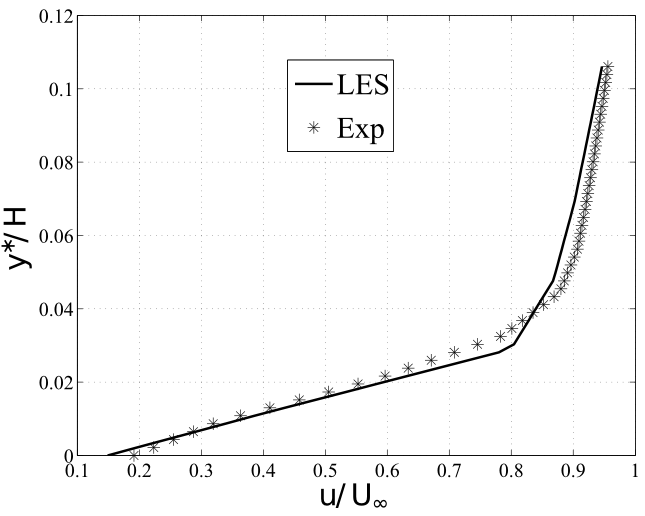}(\textit{b})\includegraphics[height=50mm,trim=0cm 0cm 0cm 0cm, clip=true, keepaspectratio]{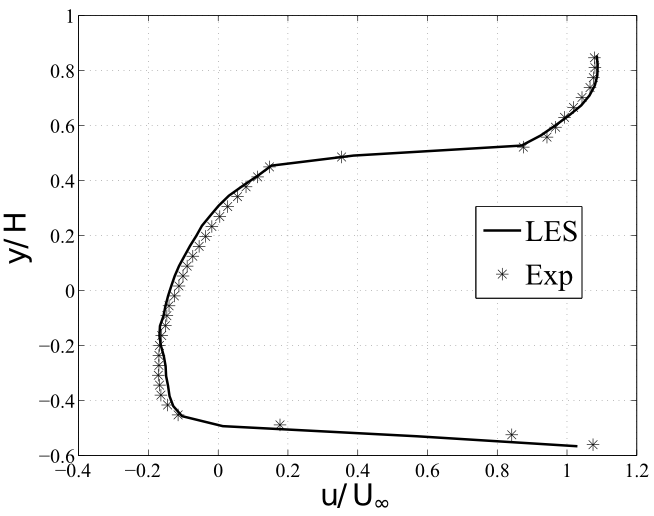}}
\centerline{(\textit{c})\includegraphics[height=50mm,trim=0cm 0cm 0cm 0cm, clip=true, keepaspectratio]{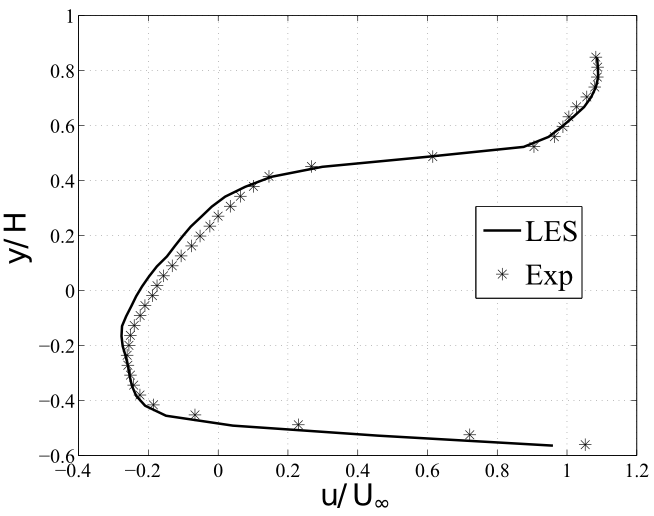}(\textit{d})\includegraphics[height=50mm,trim=0cm 0cm 0cm 0cm, clip=true, keepaspectratio]{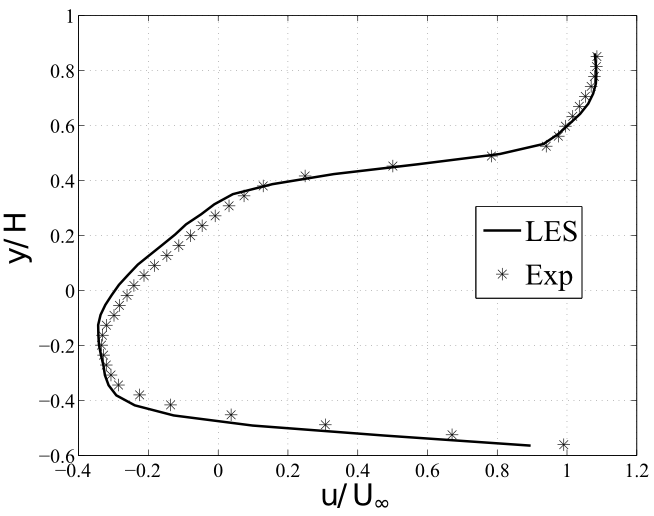}}
\centerline{(\textit{e})\includegraphics[height=50mm,trim=0cm 0cm 0cm 0cm, clip=true, keepaspectratio]{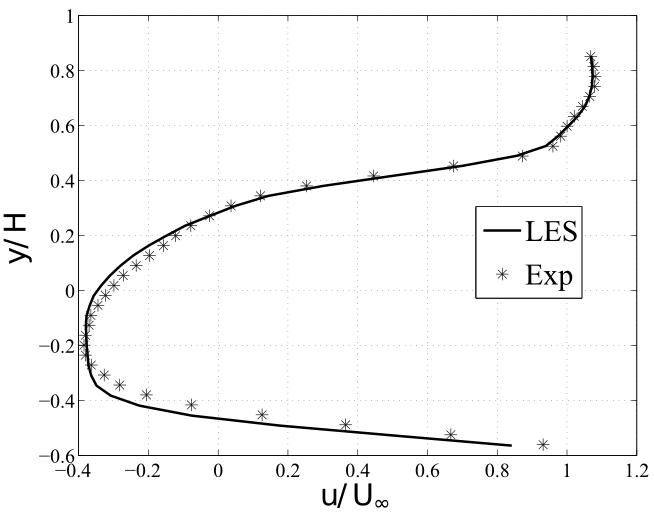}(\textit{f})\includegraphics[height=50mm,trim=0cm 0cm 0cm 0cm, clip=true, keepaspectratio]{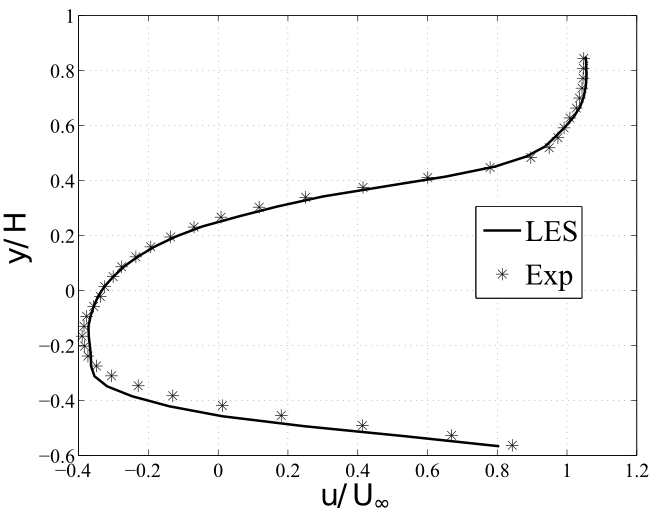}}
\caption{Profiles showing comparison of the time-averaged streamwise velocity component, $u$, for different locations in the wake. (\textit{a}) Shear layer profile $0.03\,H$ downstream of the top trailing edge. $y^* = y + 0.5\,H$. (\textit{b}) $0.17\,H$ downstream; (\textit{c}) $0.34\,H$ downstream; (\textit{d}) $0.5\,H$ downstream; (\textit{e}) $0.67\,H$ downstream; (\textit{f}) $0.84\,H$ downstream.}
\label{fig:profile}
\end{figure}

Figure \ref{fig:profile} presents a comparison between the LES results and the experimental data for the time-averaged streamwise velocity component, $u$, at different location in the symmetry plane of the wake.
The shear layer profile slightly downstream ($0.03\,H$) of the top trailing edge of the PIV data and the LES data are presented in figure \ref{fig:profile}(\textit{a}). The two profiles are in good agreement. The slow recovery of the shear layer is due to momentum loss in the separation on the front edges of the body. Such slow recovery of the shear layer profile at the trailing edge was found in the experimental study by \cite{Grandemange2013jfm} likewise.
Figures \ref{fig:profile}(\textit{b})-(\textit{f}) present profiles along lines extending from the ground to a position above the wake at five different streamwise locations in the wake. None of the profiles shows any significant discrepancy between the LES and the PIV data.

\subsection{Streamlines of time-averaged velocity in symmetry plane}
\begin{figure}
(\textit{a})\includegraphics[height=40mm,trim=2cm 0cm 2cm 0cm, clip=true, keepaspectratio]{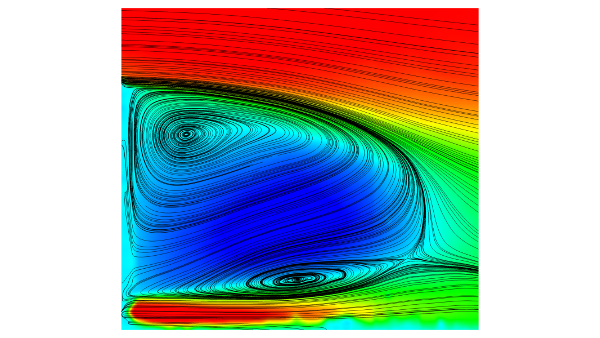}(\textit{b})\includegraphics[height=40mm,trim=2cm 0cm 2cm 0cm, clip=true, keepaspectratio]{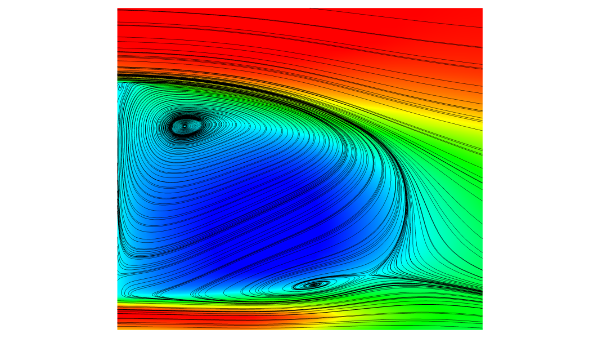} \includegraphics[height=40mm,trim=8cm 6cm 7cm 6cm,clip=true,keepaspectratio]{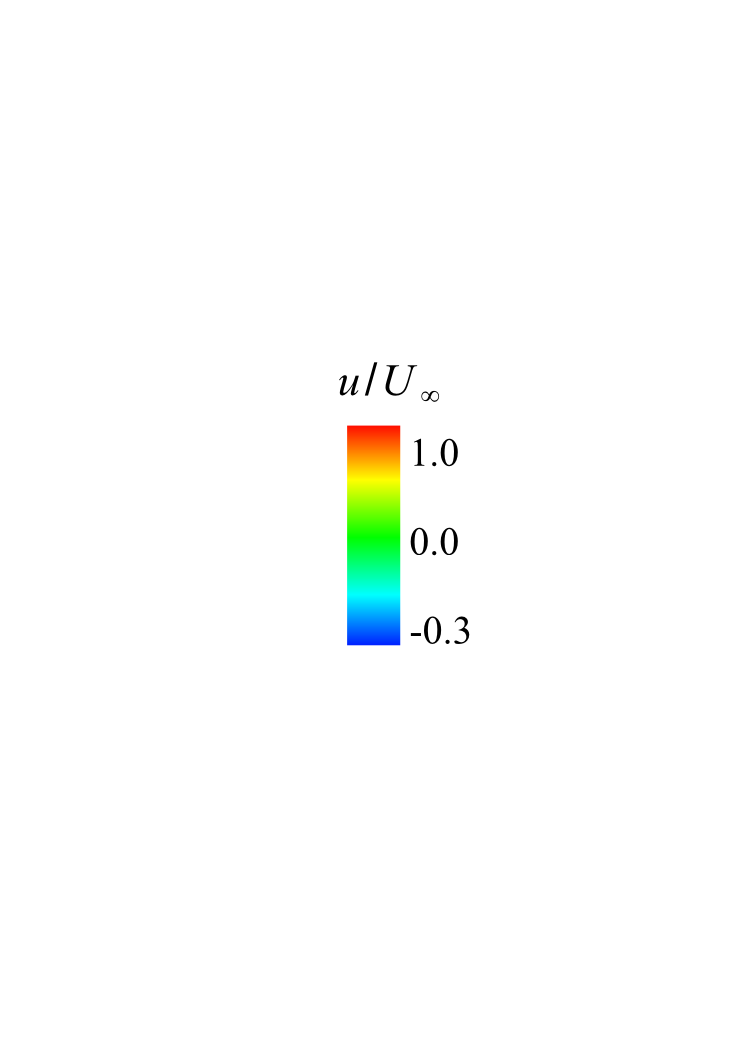}
(\textit{c})\includegraphics[height=40mm,trim=2cm 0cm 2cm 0cm, clip=true, keepaspectratio]{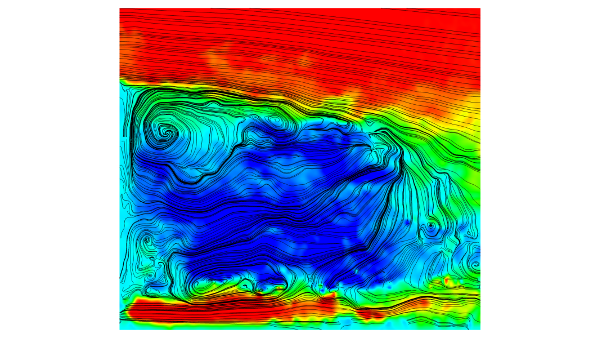}(\textit{d})\includegraphics[height=40mm,trim=2cm 0cm 2cm 0cm, clip=true, keepaspectratio]{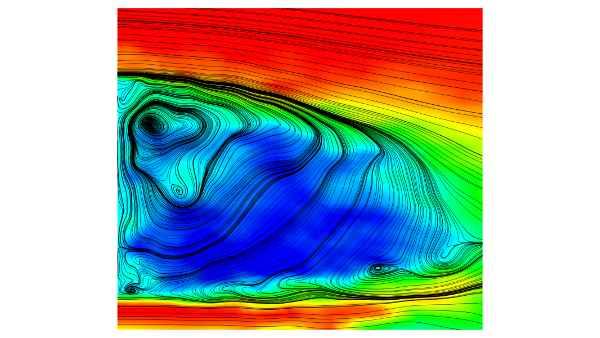}
\caption{Comparison between experimental PIV data and LES data in the symmetry plane. (\textit{a}) time-averaged PIV; (\textit{b}) time-averaged LES. (\textit{c}) one instantaneous realization of PIV data; (\textit{d}) one instantaneous realization of LES data. The length of the planes is $1.7\,H$ and the height is $1.5\,H$.}
\label{fig:streamscomp}
\end{figure}

Figures \ref{fig:streamscomp}(\textit{a}) and (\textit{b}) present the time-averaged flow in the symmetry plane from the PIV data and the LES simulation. The upper center of the time-averaged toro\"{\i}dal vortical structure in the wake is located closer to base than the lower center. The organisation of the flow in the wake is very sensitive to the set-up, in particular the gap clearance between the body and the ground. Therefore, similar studies of the geometry show different organisation of the wake. In the study by \cite{Grandemange2013jfm}, the location of the upper vortical center is located further downstream, at the same distance from the base as the lower center. However, the gap width was less in that study than in the present study, and the Reynolds number as well. Figures \ref{fig:streamscomp}(\textit{c}) and (\textit{d}) shows one instantaneous realization from the PIV and LES data, respectively.

\subsection{Spectra of transversal velocity component}
\begin{figure}
\centerline{\includegraphics[height=50mm,trim=0cm 0cm 0cm 0cm, clip=true, keepaspectratio]{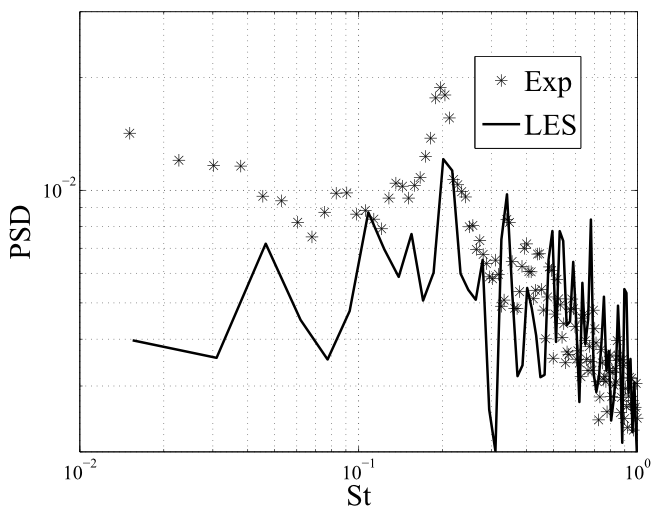}}
\caption{Power Spectral Density (PSD) of the transversal velocity component, $v$, at the point $x = 2.25\,H$, $y = 0.34\,H$, of LES data and of the velocity magnitude of the hot wire data.}
\label{fig:spectra}
\end{figure}

Spectra of velocity are presented in figure \ref{fig:spectra} at a point located downstream of the separation region. Both the hotwire data and the LES data show a signature at $St \approx$ 0.2, corresponding to the global shedding of the wake. This peak was also found in the study by \cite{LahayeIJA2014}.

\section{\textcolor{black}{Spatial resolution in LES simulation}}
\label{appB}
\begin{table}
  \begin{center}
\def~{\hphantom{0}}
  \begin{tabular}{ccc}
 $mean(\Delta n^+)$ & $mean(\Delta x^+)$ & $mean(\Delta s^+)$ \\\\[3pt]
 0.55 & 80 & 20 \\
  \end{tabular}
  \caption{Spatial resolution in the LES simulation.}
  \label{tab:spatialresolution}
  \end{center}
\end{table}

The time-averaged spatial resolution on the body expressed in viscous wall units, $\Delta n^+ = \Delta n / \lambda^+,$ $\Delta x^+ = \Delta x / \lambda^+$ and $\Delta s^+ = \Delta s / \lambda^+$, are presented in table \ref{tab:spatialresolution}. $\Delta n$, $\Delta x$ and $\Delta s$ refer to the sizes of the cells in the wall-normal direction, streamwise direction and the spanwise directions, respectively. $\lambda^+$ is the viscous length scale defined as $\lambda^+ = \nu /  u^*$, where $u^*$ is the wall friction velocity. The size of the cells in normal direction on the body, $n^+$, is everywhere less than 1. The spatial- and time-average of the viscous length scale on the body, $\lambda ^+$, was computed to be $0.0002\,H$. The values presented in table \ref{tab:spatialresolution} refer to the mean values on the body.
\begin{figure}
\centerline{\includegraphics[height=50mm,trim=3cm 10cm 1cm 10cm, clip=true, keepaspectratio]{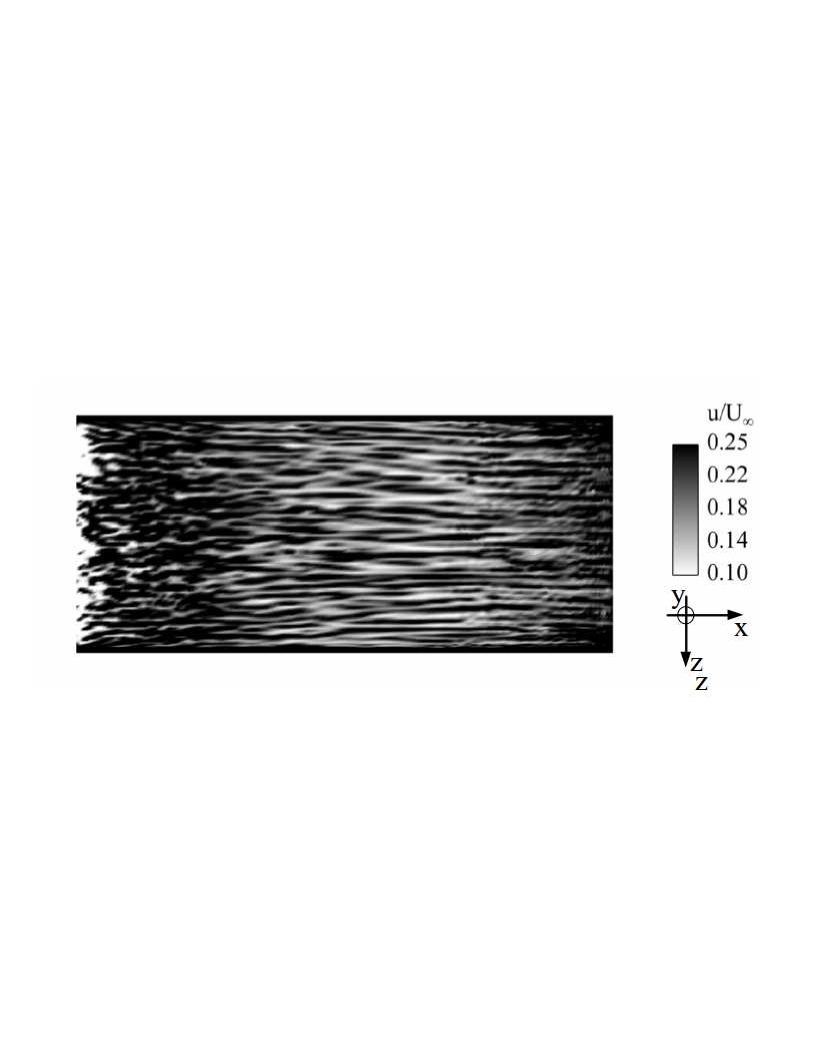}}
\caption{A plane cut from the LES simulation in the fifth cell layer away from the roof at approximately $y/\lambda ^+ \approx 5$ ,showing regions of low and high speed streamwise velocities. The length of the plane is approximately 13000$\lambda^+$ ($2.7\,H$).}
\label{fig:streaks_roof}
\end{figure}

Figure \ref{fig:streaks_roof} shows the streamwise velocity component in a plane cut in the inner boundary layer on the roof. The figure reveals low and high speed streaks in the streamwise direction, indicating the high spatial resolution in the LES simulation.


\end{document}